%% file: main.tex
\documentclass[acmsmall,screen,authorversion,nonacm,review=false,timestamp=false]{acmart}

\AtBeginDocument{%
  \providecommand\BibTeX{{%
    \normalfont B\kern-0.5em{\scshape i\kern-0.25em b}\kern-0.8em\TeX}}}

\usepackage{caption}
\usepackage{subcaption}
\begin{document}

\title{Reducing Minor Page Fault Overheads through Enhanced Page Walker} 


\author{Chandrahas Tirumalasetty}
\email{chandrhas996@tamu.edu}
\affiliation{
\institution{Department of Electrical \& Computer Engineering, TAMU}
\city{College Station}
\state{Texas}
\country{USA}
}
\author{Chih Chieh Chou}
\email{ccchou2003@tamu.edu}
\affiliation{
\institution{Amazon Lab 126}
\state{California}
\country{USA}
}
\author{Narasimha Reddy}
\email{reddy@tamu.edu}
\affiliation{
\institution{Department of Electrical \& Computer Engineering, TAMU}
\city{College Station}
\state{Texas}
\country{USA}
}
\author{Paul Gratz}
\email{pgratz@gratz1.com}
\affiliation{
\institution{Department of Electrical \& Computer Engineering, TAMU}
\city{College Station}
\state{Texas}
\country{USA}
}
\author{Ayman Abouelwafa}
\email{ayman.abouelwafa@hpe.com}
\affiliation{
\institution{Hewlett Packard Enterprise}
\country{USA}
}


\input{abstract.tex}
\maketitle


\input{intro.tex}

\input{background.tex}
\input{design.tex}

\input{impl.tex}

\input{eval.tex}

\input{conclusion.tex}
\bibliographystyle{ACM-Reference-Format}
\bibliography{references}
\end{document}

%% file: abstract.tex
\begin{abstract}
\label{sec:Abstract}

Application virtual memory footprints are growing rapidly in all systems from servers down to smartphones.  To address this growing demand, system integrators are incorporating ever larger amounts of main memory, warranting rethinking of memory management.  In current systems, applications produce page fault exceptions whenever they access virtual memory regions which are not backed by a physical page.  As application memory footprints grow, they induce more and more minor pagefaults. 
Handling of each minor page fault can take few 1000’s of CPU-cycles and blocks the application till OS kernel finds a free physical frame. These page faults can be detrimental to the performance, when their frequency of occurrence is high and spread across application run-time.  Specifically, lazy allocation induced minor page faults are increasingly impacting application performance.  Our evaluation of several workloads indicates an overhead due to minor page faults as high as 29\% of execution time.

In this paper, we propose to mitigate this problem through a hardware, software co-design approach.  Specifically we first propose to parallelize portions of the kernel page allocation to run ahead of fault time in a separate thread.  Then we propose the Minor Fault Offload Engine(MFOE), a per-core hardware accelerator for minor fault handling. MFOE is equipped with pre-allocated page frame table that it uses to service a page fault.  On a page fault, MFOE quickly picks a pre-allocated page frame from this table, makes an entry for it in the TLB, and updates the page table entry to satisfy the page fault. The pre-allocation frame tables are periodically refreshed by a background kernel thread, which also updates the data structures in the kernel to account for the handled page faults. We evaluate this system in the gem5 architectural simulator with a modified Linux kernel running on top of simulated hardware containing the MFOE accelerator.  Our results show that MFOE improves the average critical-path fault handling latency by 33$\times$ and tail critical path latency by 51$\times$. Amongst the evaluated applications, we observed an improvement of run-time by an average of 6.6\%.

\end{abstract}

%% file: intro.tex
\section{Introduction}
\label{sec:MFOEintroduction}

Virtual memory is a useful and critical mechanism used in many modern operating systems which helps in managing memory resources while giving applications the illusion of owning the entire (or sometimes more than) available physical memory.
To do so, virtual memory basically relies on two fundamental techniques:
(1) Lazy allocation and (2) Swapping.

Lazy allocation is a policy of physical memory allocation wherein memory pages are actually allocated only when accessed.
By doing so, the OS memory subsystem does not have to allocate applications any memory at \texttt{malloc} time; instead, memory is allocated when memory is indeed needed, upon its first use.
Lazy allocation can work because it is very rare for an application to touch/access all pages it requests immediately after application requests it.
Typically, program working sets are much smaller than their whole memory footprints.
Based on this characteristic of programs, the memory subsystem can let multiple applications concurrently execute and therefore improve the overall system performance without creating a scarcity of memory. Even as lazy allocation reduces the actual allocation of memory to applications, it is still possible that all memory becomes allocated when many applications are executing concurrently in the system. If the memory runs out, the swapping mechanism is adopted to store the content of some memory pages in the non-volatile storage devices and those memory pages, after their contents are wiped out, are reused and re-allocated to other applications.

These two mechanisms usually work as part of the kernel page fault exception handling.
When the memory is "allocated" (via \texttt{mmap} or \texttt{malloc}), 
actually only a region of the virtual address space is created by the kernel for the calling program; physical pages are not allocated to pair with those virtual pages until they are touched.
A memory access (touch) within this newly created region triggers a page fault exception by the hardware page walker, and then the kernel (software) exception handler will check and confirm 
this access as legal and in turn allocates a physical page for it. 

The page fault exception handler must carry out a number of operations to handle a page fault. Some of these operations include checking validity of the faulting address, acquiring a page from available free memory page pool, filling the page with the corresponding data content from storage device (major page fault) or zeroing the page (minor page fault), creating some data structures for memory management.
These operations result in considerable overhead for handling page faults. As we will show later, page fault handling can result in overheads of thousands of processor cycles. 

This overhead was not a big concern when pages are accessed from slow devices. With faster SSDs and Non Volatile Memory (NVM) storage devices, this overhead has recently received attention \cite{sigarch20}. 
Our work here focuses on the impact of this overhead on minor page faults. We show that even when sufficient memory is available, overheads from handling minor page faults can contribute to considerable performance loss.  Table~\ref{table:overhead} shows that the overhead from handling minor page faults can range up to 29\% of the application run time. The overhead calculated is just the measure of time spent in the kernel to handle minor page faults. But actual overhead due to minor faults also includes page table walks and context switching, and is expected to be slightly higher than the fraction shown.


We highlight the impact of minor page faults on one of the popular applications, Function-as-a-Service (FaaS). The latency due to minor page-fault handling is a non-trivial contributor to invocation latency in function-as-service(FaaS) systems. More specifically, we observed that minor fault frequency increases with number of cold starts. We use the faas-profiler framework of ~\cite{10.1145/3352460.3358296}, to measure the program invocation rate and the minor fault rate. For this experiment, we have chosen OCR-img application which needs a container run-time. As shown in Fig.\ref{fig:faas_experiment}, for the given application, the minor fault rate reaches the peak at around 110 Kfaults/sec for 20 invocations/sec. As we increase the invocation rate beyond 20 invocations/sec, the minor-fault rate drops off. This happens because we have exceeded the system capacity for invocations at 20 invocations/sec for this particular application and each invocation's latency at this stage is more dependent on wait time of the invocation inside queue. 
FaaS providers charge their users based on the latency of invocations. The system spending more time in the kernel handling minor faults results in users paying the bill for non-application run-time.

More discussion on benchmarks and the measurement methodology is provided in Section ~\ref{subsec:methodology}.

 \begin{minipage}{\textwidth}
  \begin{minipage}[b]{0.5\textwidth}
    \begin{tabular}{|c|c|c|} 
    \hline
    \textbf{Benchmark} &\textbf{Avg.} &\textbf{Fault}\\
    & \textbf{Fault rate} & \textbf{Overhead} \\
    & \textbf{(K.faults/sec)} & (\%)\\\hline
    GCC compiler & 365.31 & 29.09 \\ \hline
   FaaS & 108.92 & 9.05 \\ \hline
    Parsec-dedup & 165.88 & 8.26 \\ \hline
    YCSB-memcached & 120.06 & 6.47\\ \hline
    Splash2X-radix & 259.25 & 16.17\\ \hline
    Splash2x-fft & 222.62 & 10.06\\ \hline
    XSBench & 89.09 & 4.90 \\ \hline
    GAP-BC & 77.74 & 3.08 \\ \hline
    Integer Sort & 104.91 & 4.20 \\ \hline
\end{tabular}
\label{table:overhead}
\captionof{table}{Benchmarks with anonymous page fault overhead}
\end{minipage}
\begin{minipage}[b]{0.55\textwidth}
\centering
\includegraphics[scale=0.425]{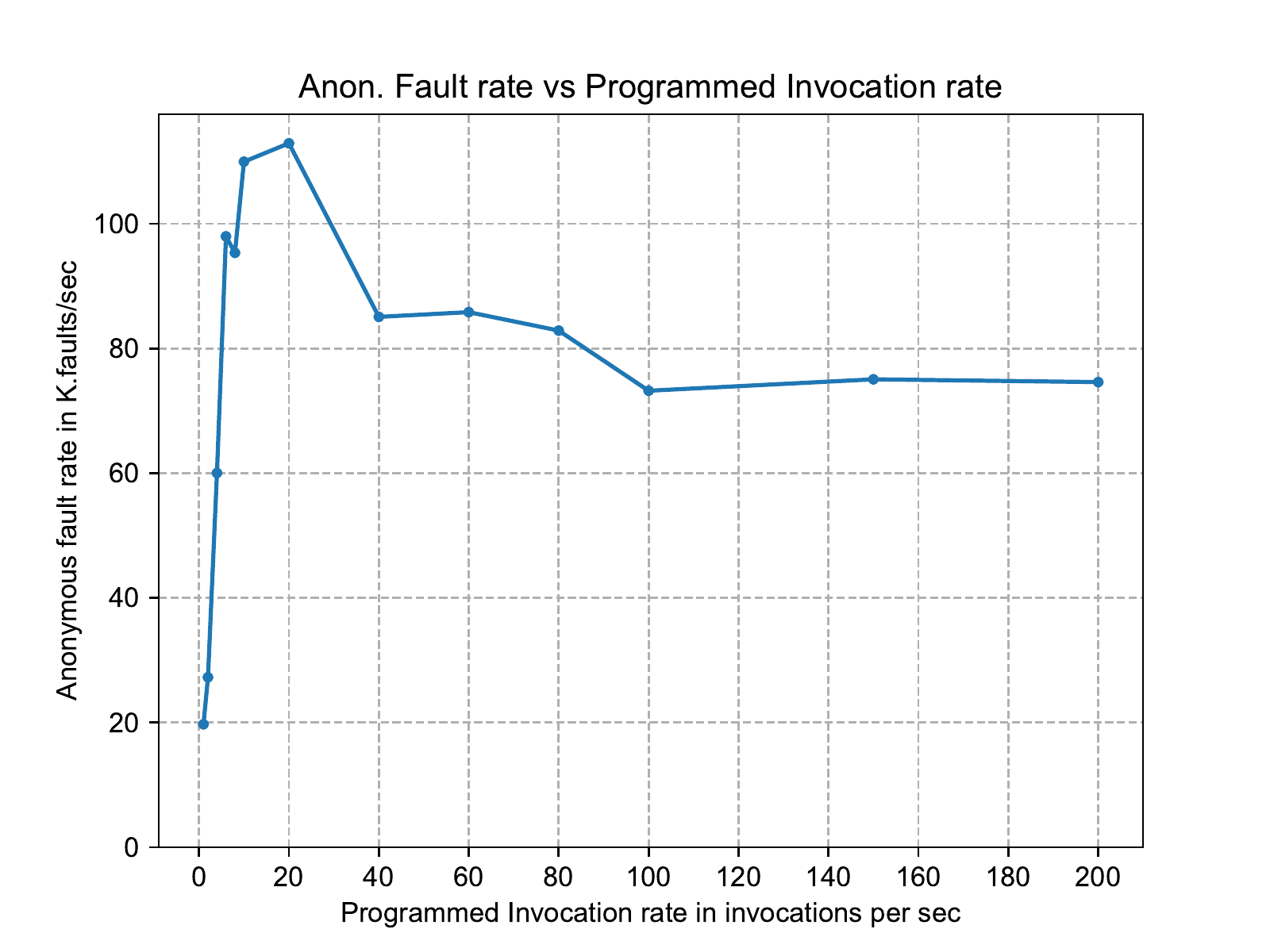}
\captionof{figure}{Minor-Fault rate vs programmed \newline invocation rate}
\label{fig:faas_experiment}
\end{minipage}
\end{minipage}



Here we propose a new approach to memory management, with the goal of minimizing overheads for minor page faults in such applications. 
Our proposed approach rethinks how memory management function can be organized and realigns the work to be done within the OS and hardware. 
We move majority of the work in handling a minor page fault to the background through pre-allocation of pages, instead of current model of on-demand inline allocation of pages to faulting processes. Our approach carries out minor page fault handling within the context of the faulting process through hardware, thus obviating the context switch of traditional approaches in current OSs.  This approach moves most of the OS work from the critical path of the application's execution to a separate, ahead of time thread, with the remainder being accelerated in hardware upon minor fault demand. 
This requires rearranging the memory management functions in the kernel space of the OS and redesigning the hardware page walker to carry out the remaining minimal processing in hardware at the time of a page fault, within the context of the faulting process.



The contributions of this paper are as follows:
\begin{itemize}
\item Proposes a new page fault handling architecture, composed of a coordinated kernel thread and hardware page walker, to reduce the inline page fault overheads.

\item Proposes a redesign of memory management functions in the OS and hardware to enable the realization of the proposed architecture.

\item Implements the proposed changes to processor architecture on gem5 simulator and modifies the existing Linux kernel to
accomplish our new minor page fault handling mechanism.

\item Evaluates the proposed minor page fault handling mechanism using microbenchmarks to show significant critical path latency improvements.

\item Shows that the proposed mechanisms can improve full application running times by 1\%-25.4\%. 
\vspace{-0.5em}
\end{itemize}

The remainder of this paper is organized as follows.
Section~\ref{sec:POEbackground} describes the background and related
work.
Section~\ref{sec:POEdesign} presents the motivations and key design concepts
of software and hardware modifications of our optimized page fault handling.
Section~\ref{sec:MFOEimplementation} explains the
implementation of our new page fault handling approach by modifying the Linux kernel and x86-64 CPUs on gem5 simulator in more detail.  
Section~\ref{sec:POEevaluation} presents our results of
evaluation with benchmarks.  
Finally, section~\ref{sec:POEconclusion} provides a conclusion.

%% file: background.tex
\section{Background}
\label{sec:POEbackground}



\subsection{Page Fault}
User programs request memory using \texttt{malloc} system call or \texttt{mmap}. However both of these calls, do not allocate any memory for applications when called. Instead, they only create and validate a region of virtual address space for the calling process and return the start address of this region. Later, when applications access some address within this newly created region (if no page has been allocated for this address yet) an exception(page fault) would be triggered by hardware. Execution context is changed from userspace to the kernel and the page fault handler of the kernel starts running on behalf of the faulting application. The handler allocates a page for the faulting address, updates the page-table entries and updates the kernel book-keeping data structures.  This kind of fault is called a minor page fault.

After the page fault exception handler completes its job, the faulting load/store instruction is re-executed. This mode changing (from user mode to kernel mode and then back to user mode) requires some ``state'', such as local variables, hardware registers, program counter, etc.) of the user program to be stored (or pushed) on the stack. The last few operations of the exception handler is to pop that state from stack back to original place, so that the faulting instruction can continue to execute.
In addition to the overhead of pushing onto and popping from stack, context switch results in pollution of architectural resources like caches, TLB, branch predictors etc. 

In systems with non-volatile memory(NVM) devices, memory can be accessed by applications through the file system read/write interface. This method does not incur page faults, but it needs to rely on system calls, which can also result in context switching. A large body of recent work has observed that systems software overhead incurred by file system APIs is high relative to NVM access times and tries to reduce this overhead through various methods 
\cite{strata, scmfs, vnvml,breeze,MOD}.

\vspace{-1em}
\subsection{Hardware Page  Walker}

Hardware page walker \cite{walker} is popularly employed in modern CPUs. On a TLB miss in x86\_64 architecture, the hardware page walker assumes the job of reading the faulting address from CR3 register and "walk" the multiple levels of the page table to find the appropriate page table entry (PTE).  If the page walker can reach the lowest PTE of the faulting address and finds the present bit of PTE is set; meaning that a page has been allocated for this virtual address and its physical frame number is also stored at the PTE, then the page walker would simply update the corresponding TLB entry and re-execute the faulting instruction again. 
Otherwise, the page walker will trigger a page fault exception and let the kernel handle the page fault.

Enhancements to page walkers have been proposed \cite{pagewalker1,PW-mod1} previously to achieve different goals. Our work here can be viewed as an enhancement to the hardware page walker to carry out the entirety of page fault handling's critical path.

\vspace{-1em}
\subsection{Related Work}
\subsubsection{Hardware Paging}
Recent work \cite{sigarch20} advocated hardware based demand paging for major page faults. This work has similar motivation as our work i.e. reducing the cost of page faults, but focuses on major page faults. Our work here shows the importance of handling minor page faults even when the data sets can be entirely memory resident. Nevertheless, our approach could be extended to support major faults. In contrast to a single pre-allocation queue proposed in earlier work, we have proposed pre-allocation tables for each core, thus providing flexibility to implement NUMA policies on pre-allocated pages.  Major page-fault handling latency was observed to be an overhead when paging is used as a mechanism to fetch pages from a disaggregated memory node \cite{pemberton2018enabling}. Their approach is similar to ours i.e. they have offloaded fault handling to HW accompanied by kernel changes, but targets different systems.

Adding page-fault support to network controllers to avoid page pinning for DMA requests is proposed in \cite{10.1145/3037697.3037710}. Our page-fault handling mechanism is orthogonal to this proposal and can be integrated with this work, to enhance the fault handling latency.
Hardware based page-fault handler is used  for virtual memory implementation at disaggregated memory node in \cite{guo2021clio}. This proposal focuses on implementing light weight virtual memory for remote memory accesses with offloading of fast path to ASIC \& FPGA, and slow path to an ARM processor, in a co-designed memory node. Albeit the design similarities, this proposal is unrelated to our approach in this paper. Our approach focuses on offloading generic anonymous fault handling to an enhanced page table walker.

\subsubsection{TLB \& Paging Optimizations}
Large pages have been proposed to reduce the costs of number of page faults to be handled \cite{LargePage1,LargePage2,range-translations,LargePage-Ingens} and sometimes to increase the reach of the TLBs \cite{TLB-extend1}. Large pages are beneficial when addresses are allocated over contiguous ranges, but many new applications allocate small objects \cite{FB-workload}, and have required additional support for efficiencies \cite{perforated-pages}. From a memory management perspective, the kernel has the job of promoting and demoting huge pages based on their usage. Also the kernel has to perform memory compaction of base(4K) pages, to ensure that it has free 2MB pages for future allocation request. 
Our approach here works across different workloads and moves much of the memory management overhead away from inline processes. 
Mitosis\cite{10.1145/3373376.3378468} replicates page table entries in multi-node systems, to avoid costly remote accesses during page table walks by HW page table walker. Our approach makes use of the hardware page table walker, but doesn't put any restrictions on page table entry placement. Hence, MFOE can be made to work with Mitosis. 

Recent work \cite{10.1109/ISCA52012.2021.00047} uses intermediate address space-Midgard, to manage the virtual memory of applications.  Midgard  uses virtual memory areas (VMA's) instead of physical pages for memory management. Since an application has a small number of VMA's to track, it achieves low address translation overhead. Although, Midgard increases the efficiency of TLB, it doesn't remove the cost of page faults altogether. So, the overhead due to page faults exists even in systems that use Midgard.

\subsubsection{Miscellaneous}
Tracking PTE access bits to avoid unnecessary TLB shoot-downs has been proposed in \cite{203151}.  Our enhanced hardware page table walker, sets the PTE access bit during both page walk, and page fault handling in hardware. Hence no disruption to the page access tracking mechanism is caused by our enhanced page table walker.
Prefetching has also been proposed to reduce the cost of misses on page accesses \cite{SPAN, Translation-prefetching}. Prefetching is complementary and can work with our approach.
Alam et al. \cite{DIY} adopt a hardware helper thread to reduce the number of context switches incurred by page faults.  Their work, however, requires a pair of registers to indicate a single region of virtual address space (one for the start address and the other for the end address).  Their approach would be better for virtual machine workloads because a guest OS will allocate a huge amount of contiguous virtual memory region from the hypervisor as its physical memory, but might not be suitable for general workloads.

%% file: design.tex
\section{Design Overview}
\label{sec:POEdesign}

In this section we  discuss the design choices that form the basis for our enhanced page fault handling, through Minor Fault Offload Engine(MFOE). While the discussion is in the context of Intel X86\_64 and Linux OS, our modified fault handler could be implemented in any other architecture and any other operating system.

We have broken down the page fault handling in software into 3 parts- tasks that can be done before the page fault, tasks to be done for handling the page fault(critical path) and tasks that can be done after the page fault. Design of MFOE is primarily motivated by two main observations pertaining to these sub-tasks.(1) Tasks that can be done before and after the page fault occurrence, can be removed from the critical path to reduce the fault handling latency. (2) The critical path tasks are hardware amenable, hence would benefit if the hardware(MFOE) executes the critical path instead of software.

\begin{figure}[h]
\flushleft
\begin{subfigure}[b]{0.49\textwidth}
\includegraphics[scale=0.24]{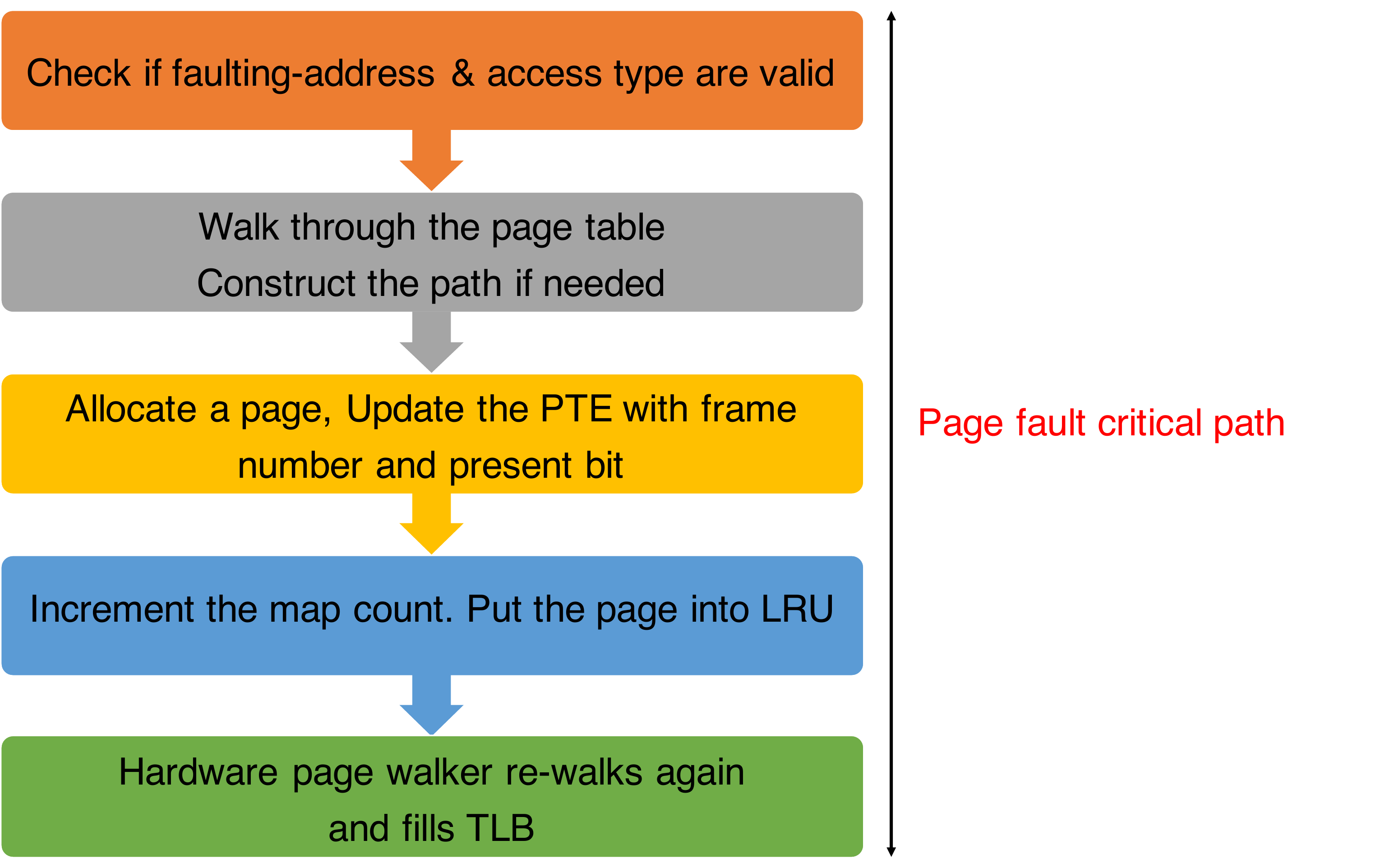}
\caption{Typical minor page-fault handling flow}
\end{subfigure}
\hfill
\begin{subfigure}[b]{0.49\textwidth}
\flushright
\includegraphics[scale=0.24]{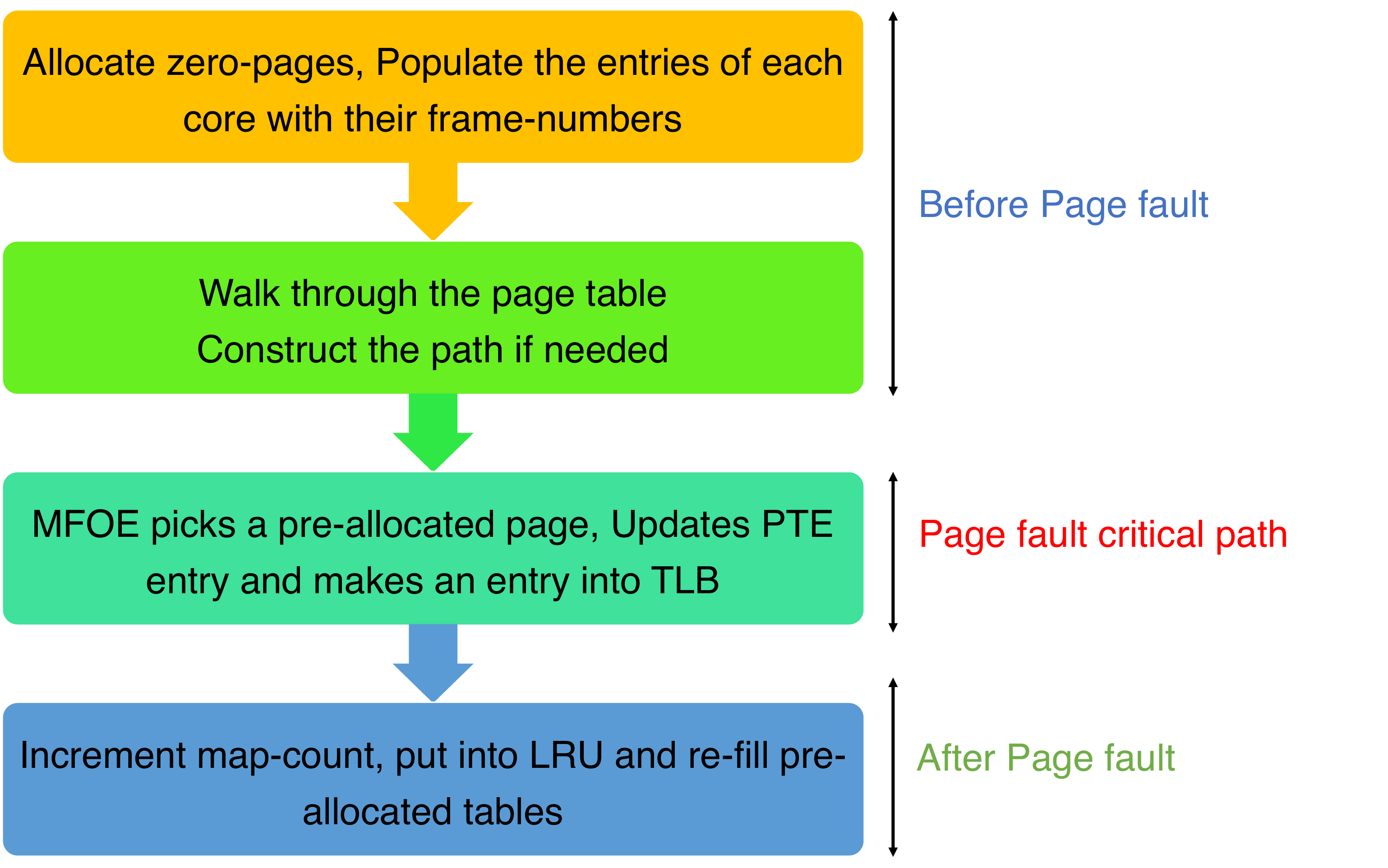}
\caption{Minor page-fault handling flow with MFOE}
\end{subfigure}
\caption{Default and MFOE fault handling methods}
\label{pagefaultflow}
\vspace{-2em}
\end{figure}

Fig.\ref{pagefaultflow} contrasts the page fault handling flow of MFOE with typical page fault handling in software. We categorize physical page allocation and constructing the page table as pre-fault tasks. Post fault tasks include updating the kernel metadata structures and establishing the reverse mappings for handled page fault. For the critical path, the MFOE has to check the validity of the faulting address, update the page table entry by consuming the physical page made available by the pre-fault task and lastly make information available for post fault maintenance routine. Pre-fault and post-fault functions are combined into a background thread, so their latencies are removed from the fault handing function in the context of the faulting program. 

This section is organized as follows, we discuss the page pre-allocation and legal virtual address space indication in \ref{subsec:MFOEpagepreallocation} and \ref{subsec:MFOEPreallocationBit} respectively, Both of these are pre-fault tasks that are moved into background. In \ref{subsec:MFOEPageTableWalker} we discuss the design of MFOE and how it handles page fault by working in conjunction with the background thread. In \ref{subsec:MFOEpostpagefaulthandling} we discuss post-fault handling tasks.


\subsection{Page Pre-Allocation}
\label{subsec:MFOEpagepreallocation}
The existing page fault handler obtains a physical page from the buddy system\footnote{ The buddy system is an organizational data-structure used by the Linux kernel to manage available physical memory frames/pages. The available memory is broken into groups of blocks, each of size power of two of base page size(4K). Based on the incoming request for free pages, the blocks are broken down into two blocks half the original size. Each broken down block being "buddies" to each other\cite{buddyallocator}}; this action may result in blocking the faulting program if no page is available and kernel must try hard to reclaim pages either from the page cache, or by writing some dirty pages to swap space as a last resort. The latter action would require issuing I/O requests to the backing storage. So its reasonable not to attempt offload such operations to hardware. Nevertheless, physical page allocation can be done without the knowledge of the faulting virtual address, hence can run before hand in background i.e. pre-allocation. We consider a mechanism to provide pre-allocated pages, these pages are used by hardware when page fault handing is completely done by hardware or by critical-path software when MFOE is employed only as a software solution.

When a page is allocated, usually a pointer of  \texttt{struct page}  (or the virtual address in kernel of that page) is returned by the buddy system in the kernel. This pointer (of \texttt{struct page}) can be translated into physical page frame number of that page easily by \texttt{page\_to\_pfn} macro.  If those frame numbers of pre-allocated pages can be saved beforehand in a certain format and the hardware can access them easily when a page fault occurs, we will be able to use a free page frame to service the fault as well as remove page allocation out of the critical path.

This design choice has the following benefits:
\begin{itemize}
\item  Since allocating pages might block the faulting programs, page pre-allocation by a kernel background thread only blocks this kernel thread and will not hurt (slow-down) the user application.
\item  In the existing Linux kernel, to allocate pages for \texttt{malloc} or anonymous private \texttt{mmap}, kernel always``zeros'' the entire page (4KB) or entire huge page (2MB) in the page fault critical path due to security concerns. Our pre-allocation mechanism also moves this time-consuming operation out of the page fault critical path.
\item  It reduces the inline work to be carried out at the time of a page fault and keeps the required hardware enhancements simple.
\end{itemize}
\vspace{-1em}
\subsection{Legal Virtual Address Space Indication}
\label{subsec:MFOEPreallocationBit}

The first and most important operation of the existing page fault handling routine is to determine if the faulting virtual address is legal. This check operation is implausible to be implemented in hardware. Also, the faulting virtual address can only be known at the moment of page fault occurrence and there is no way the hardware can know if the faulting address is legal, in advance. So, instead of trying to validate the faulting address in hardware, we let the software tell the hardware if the faulting address is legal address or not. 

To this end, we narrow down the scope of our page fault offloading mechanism. We limit ourselves to handling page faults happening in the user space; in particular, our new page fault mechanism currently only deals with minor page faults caused by accessing the virtual address space regions created by \texttt{malloc} and anonymous private \texttt{mmap} functions within user programs. The kernel keeps track of all the valid virtual memory areas of an application. The only missing piece is a mechanism to inform hardware of this information. The correct avenue to enable this communication would be at hardware page walker. The HW page walker will be the first to witness the faulting address in the HW. So, if the HW page walker is aware of validity of faulting address, it could revert to software handling of fault quickly. To establish this understanding, we  put an \texttt{MFOEable} bit in the page table walking path to tell the hardware that whether the currently accessed address is legal. The lowest Page Table Entry (PTE) seems to be a reasonable candidate for accommodating this indicator. Hence we re-purpose a bit in un-used(invalid) PTE to be used as a valid bit.

Therefore, our proposed flow is as follows. When applications call \texttt{malloc} or \texttt{mmap}, just before the system call is about to return, a short-lived background thread is spawned by the kernel. This background thread walks the page table path for all pages within this newly created virtual address region in the background, until all the lowest PTEs belonging to this newly created area have been reached.  If the paths to PTE of some addresses have not been constructed yet, the kernel background thread will construct these paths as the exception handler does. 
This path construction process is also moved out of the page fault critical path.
Meanwhile, this kernel background thread sets a \texttt{MFOEable} (Minor Fault Offload Enable) bit as a legal address indication for PTEs of all pages of this virtual address region.
If the page table walker does not find this bit in PTE, then it does not offload the minor fault handling to MFOE, it raises an exception and defaults to fault handling in kernel.
\subsection{MFOE-Minor Fault Offload Engine}
\label{subsec:MFOEPageTableWalker}
Our central idea is to move most of minor page fault operations out of the page fault critical path and to execute them before and after a page fault by software; only the mandatory operations are tackled by hardware during a page fault. Thus, we propose to keep MFOE as simple as possible by implementing it as an addition to the hardware page walker, TLB sub-system. 

Realizing the design considerations discussed in \ref{subsec:MFOEpagepreallocation} and \ref{subsec:MFOEPreallocationBit}, MFOE is equipped with a  pre-allocation page table whose entries contain page frame numbers of pre-allocated pages and \texttt{MFOEable} bits set at the leaf entries of page table across its legal address space.

These two components make the design of MFOE straightforward:
When a faulting address is accessed, there will always be a TLB miss because TLB only caches legal address translations. The hardware page walker starts to walk the multi-level page table until it reaches the lowest PTE.
If the page table walker finds out that the \texttt{MFOEable} bit is set (meaning the faulting address is legal), but the present bit is not set yet (meaning that this address does not have an allocated page), then the page table walker offloads the fault handling to MFOE.
MFOE picks an entry from the pre-allocation page table, updates the PTE with the page frame number of this entry as well as sets some flag bits into PTE (as software exception handler does),  creates a TLB entry, 
and continues to execute the faulting instruction without the need of kernel software. For all the other cases, different from the above-mentioned scenario, the typical kernel page fault exception is triggered by the hardware page table walker to handle the fault.

\subsection{Post-Page Fault Handling}
\label{subsec:MFOEpostpagefaulthandling}

Besides the page allocation, the address validation, and updating PTE entries, there are still some operations left  to be executed by the page fault handling routine, such as increasing the counter of \texttt{mm} object (by calling  \texttt{inc\_mm\_counter} function), Establishing the reverse map for physical page(through \texttt{page\_add\_new\_anon\_rmap} function, putting the page into LRU list for swapping later(through \texttt{lru\_cache\_add\_active\_or\_unevictable} function), and accounting for cgroup based memory usage metering (by calling \texttt{mem\_cgroup\_commit\_charge} function).
These operations can be delayed and executed later in the background, if the program is not terminated. Additionally due to pre-allocation of physical pages for future faults, in the event of program termination, there is a necessity to clean up the unused physical pages. This clean-up might not be needed in a real world implementation, if MFOE was used by default for page fault handling for all the programs.

\begin{figure}[h!]
    \centering
    \includegraphics[scale=0.4]{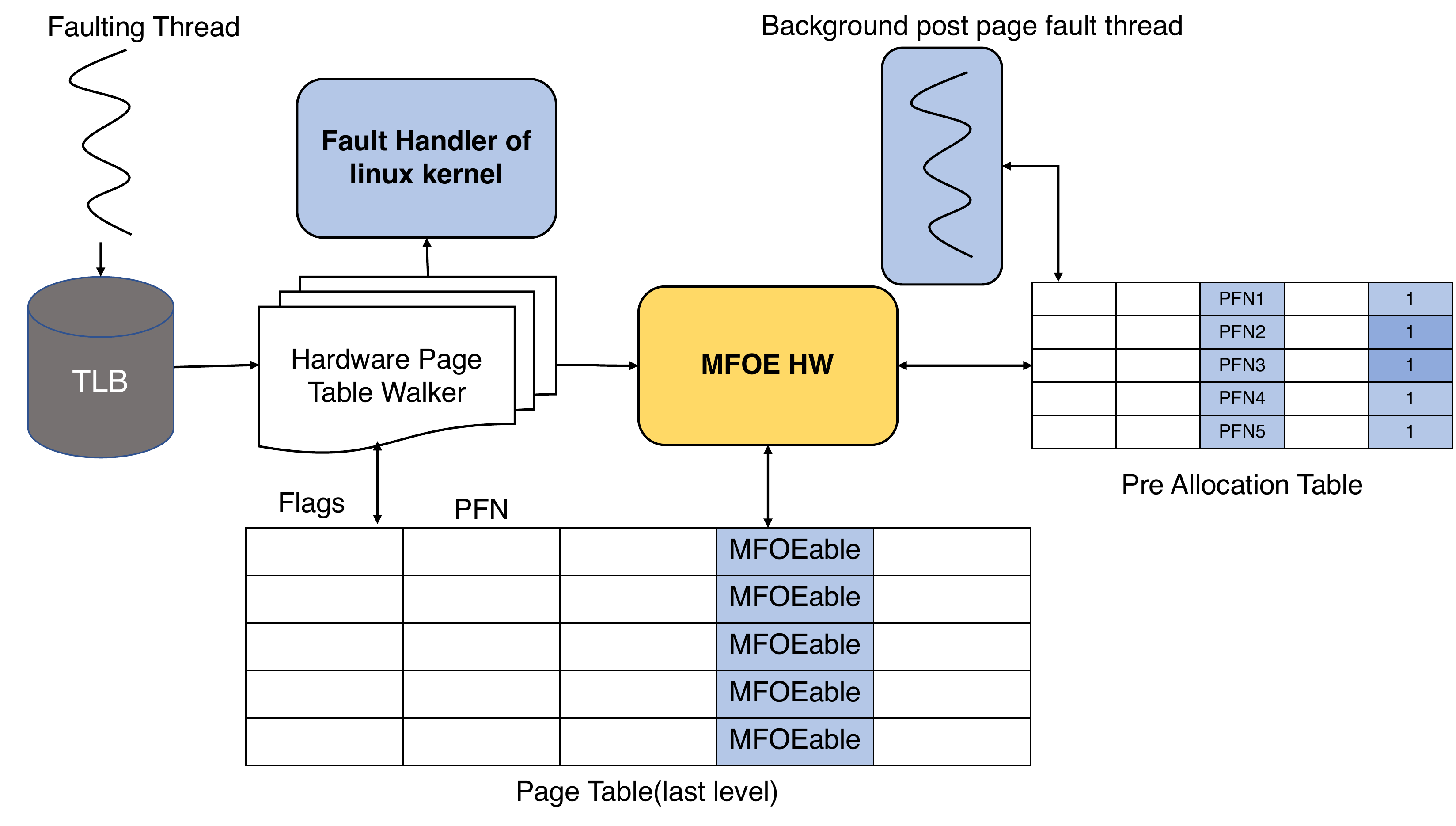}
    \vspace{-1em}
    \caption{Overall design of MFOE}
    \label{fig:mfoe_architecture}
    \vspace{-1em}
\end{figure}

Fig.\ref{fig:mfoe_architecture} details the design of MFOE and its key components.

%% file: impl.tex
\section{Implementation}
\label{sec:MFOEimplementation}

In this section, we explain the implementation details of MFOE prototype that we built. The implementation consists of both hardware(MFOE) and software(modified linux kernel) components. In \ref{subsec:MFOEenableanddisable} we discuss the system calls we implemented, to enable/disable MFOE. In \ref{subsec:MFOEpreallocationtable} we present the organization of the pre-allocation table. The details about pre-fault and post-fault components of the background thread are presented in \ref{subsec:MFOEprepagefault} \& \ref{subsec:MFOEpostpagefault} respectively. We present the details of page fault handling by hardware in \ref{subsec:MFOEpagefault}. In \ref{subsec:MFOEErrorHandling} we talk about how our implementation accounts for errors and  \ref{subsec:MFOEDisable} has a discussion on disabling MFOE.
In \ref{subsec:MFOESE}, we discuss software emulation of MFOE.Finally we analyze limitations \& issues involved with the current implementation in \ref{subsec:MFOEOtherIssues}.

\subsection{MFOE Enable/Disable System Calls}
\label{subsec:MFOEenableanddisable}

To enable optimized page fault handling through MFOE,
applications should use the \texttt{MFOE\_enable} system call at the beginning of their source code.\footnote{This is the only application level change that is required to enable MFOE}
MFOE is automatically disabled by the  \texttt{MFOE\_disable()}  function called inside \texttt{\_\_mmput()} by kernel when applications are about to be terminated. We have also implemented the \texttt{MFOE\_disable} system call, which would directly call \texttt{MFOE\_disable()} function, this allows applications to disable MFOE in the middle of their run-time if they decide to do so. If MFOE is disabled in the middle of the program run time, the PTEs that have been populated with \texttt{MFOEable} bit set will stay intact, but the faults on such pages will be defaulted to kernel page fault handling.

MFOE related software can be implemented as a kernel module and operations of  \texttt{MFOE\_enable} and \texttt{MFOE\_disable} system calls can be executed when the MFOE module is loaded and unloaded, respectively. By doing so, all user applications will automatically enable MFOE by default. However, we use a system call to simplify our evaluation of MFOE.

\subsubsection{MFOE\_enable system call}
\label{subsubsec:MFOEenable}
\texttt{MFOE\_enable} system call expects an input ``preallocation\_size'', which indicates the number of physical frames that are to be pre-allocated per core.  With ``preallocation\_size'', the kernel calculates the physical frames needed to house the per-core pre-allocation table itself. The pre-allocation table contains the frame numbers of the pre-allocated pages that can be consumed by the faults occurring on the particular core. Since our implementation chooses to have a pre-allocation table per core, we have to construct a separate table for each core. We should  mention that the construction of the pre-allocation table is not needed for every \texttt{MFOE\_enable} system call by the application, but only for the very first \texttt{MFOE\_enable} system call since the system boot. 
In our design, size of each entry in the pre-allocation table is 16 bytes. Suppose for ``preallocation\_size'' 256, we need to allocate a 4K page per core for pre-allocation table. This can be viewed as memory overhead for using MFOE.

After the pre-allocation tables are constructed, the start address of pre-allocation table is conveyed to the hardware. The \texttt{schedule\_on\_each\_cpu()} kernel function executes a small code on each core in the system. This code writes the page frame number (34 bits) of the first page of the corresponding pre-allocation table, the number of entries of the pre-allocation table (16 bits), and MFOE\_ENABLE bit (1 bit) to a newly added 64 bit register. We used the architecturally hidden CR9 control register of X86 for this purpose. 

We have made an implementation choice to construct only the pages that constitute pre-allocation table, but not the the pre-allocated physical frames that form the entries of this table. We have chosen to do so to minimize the run-time of the system call as well as the blocking time of the user-space process. A background kernel thread that is spawned at the end of system call will be responsible for populating the pre-allocation tables of all the cores. Hence the time taken to complete the pre-allocation for all the cores increases with pre-allocation width and number of cores in the system. This implementation choice is justified because we minimize the time in kernel context due to system call and avoid non-deterministic blocking time taken due to pre-allocation.

It is possible that MFOE hardware looks up the pre-allocation table to satisfy a page fault even before this aforementioned background kernel thread completes its task of filling each core's pre-allocation table. In this case, MFOE hardware fails to get a valid page frame number from the table and then will subsequently trigger a normal page fault to be handled in kernel.

To indicate to the kernel whether a process has enabled MFOE, we have added a boolean type member called ``mfoe'' to the memory descriptor \texttt{mm\_struct}. When  the program calls the \texttt{MFOE\_enable} system call , this value is set to true. The memory allocation routines(\texttt{mmap} or \texttt{malloc}) checks ``mfoe\_enable'' to see if the page table entries have to allocated differently i.e. with \texttt{MFOEable} bits.



\subsection{Pre-Allocation Table}
\label{subsec:MFOEpreallocationtable}

Our pre-allocation table adopts a lockless ring buffer architecture \cite{ringbufer} 
with one producer (the kernel background thread, which produces/pre-allocates pages) and one consumer (the MFOE hardware which consumes pages).
Each entry of pre-allocation table has 16 bytes, and the first entry (entry number 0) stands for the table header, which contains the head index (ranging from one to the number of entries), tail index (also ranging from one to the number of entries), number of table entries, and lock bits. the size of each field in header is four bytes.
Except for the table header, the entry's format is delineated in Fig.~\ref{fig:preallocationtable}. As shown below, the fields of each pre-allocation entry are faulting virtual address, TGID (thread group id), page frame number, used bit, and valid bit.

\begin{figure}[h!]
	\centering
     	\includegraphics[scale=0.4]{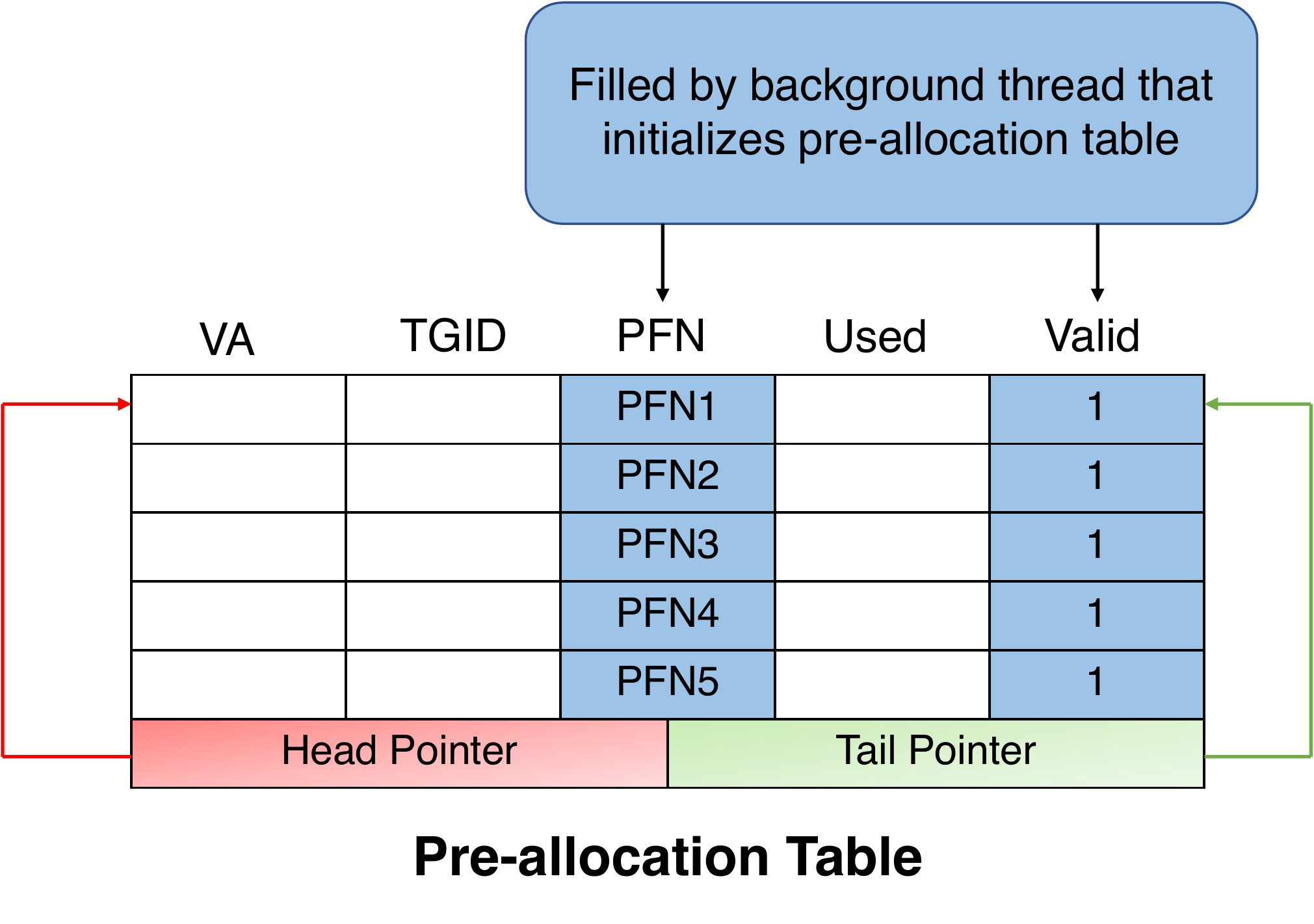}
     	\vspace{-1em}
	\caption{The pre-allocation table and its fields after pages are pre-allocated. Header not shown.}
	\label{fig:preallocationtable}
	\vspace{-1.25em}
\end{figure}

The producer only looks at the head index, and the consumer only checks the tail index. When the kernel (producer) pre-allocates a page for the pre-allocation table, first it looks up the corresponding entry of the head index from table header.
If this entry is not valid (the valid bit is not set), meaning that 
this entry does not contain a valid pre-allocated page, then the kernel allocates a page and puts its page frame number into the corresponding field of the pre-allocation entry, increments the head index by one, and checks the next entry. Kernel would continue this page pre-allocation process until it reaches an entry whose valid bit has already been set or till it traverses the entire pre-allocation table for a particular core.

Since our implementation uses a per-core pre-allocation table, the kernel, when choosing pages for pre-allocation, can follow an allocation policy. For example, if a system has multiple NUMA (non-uniform memory access) nodes,  then pages can be pre-allocated from the near memory node first.  After the pages from the near memory node are exhausted, we can decide either to pre-allocate pages from far memory node, or not to pre-allocate pages at all. Effectively with this policy, MFOE will trigger a typical page fault exception after it runs out of near memory node pages. In the current implementation we enforce a static policy where we prefer local node(to the given core) physical pages for all the pre-allocations. The kernel API used for pre-allocation(buddy-allocator) automatically falls back to remote node pages if the local node pages are unavailable. Hence, after we run out of local node pages, we will use remote node physical pages for pre-allocation.

\subsection{Pre-Page Fault Software Handling}
\label{subsec:MFOEprepagefault}
Every time an user-space application calls \texttt{malloc} or anonymous private \texttt{mmap} to allocate memory,  the kernel will check if the \texttt{mfoe} member of memory descriptor object(\texttt{mm\_struct}) is set as true. If that's the case, then kernel  will flag the virtual memory area as belonging to a MFOE enabled process. To achieve this we add a new flag \texttt{VM\_MFOE} to \texttt{struct vm\_area\_struct}\footnote{\texttt{struct vm\_area\_struct} is a data-structure used by Linux kernel to manage contiguous virtual memory areas. The memory descriptor-\texttt{mm\_struct} of each process maintains a list of \texttt{vm\_area\_struct} to represent all of process's virtual memory}. This flag will be set for the virtual memory areas belonging to the process, that are created after the process has enabled MFOE through \texttt{MFOE\_enable} system call.

At the end of \texttt{malloc} or anonymous private \texttt{mmap} function, we account for the kernel's responsibility to craft the  leaf entry of the page table with MFOEable bit. Since we know the start and end address of the newly created virtual memory area, we walk through the page table of the calling process until it reaches  all the corresponding lowest level page table entries and set the \texttt{MFOEable} bit(bit-position 2) to 1. If the walking path of the multi-level page table(non-leaf entries) has not been constructed yet,  kernel will construct it, the same way as the page fault handler does. This is the realization of design motivation provided in Section~\ref{subsec:MFOEPreallocationBit}.

\begin{figure}[h!]
	\centering
     	\includegraphics[scale=0.4]{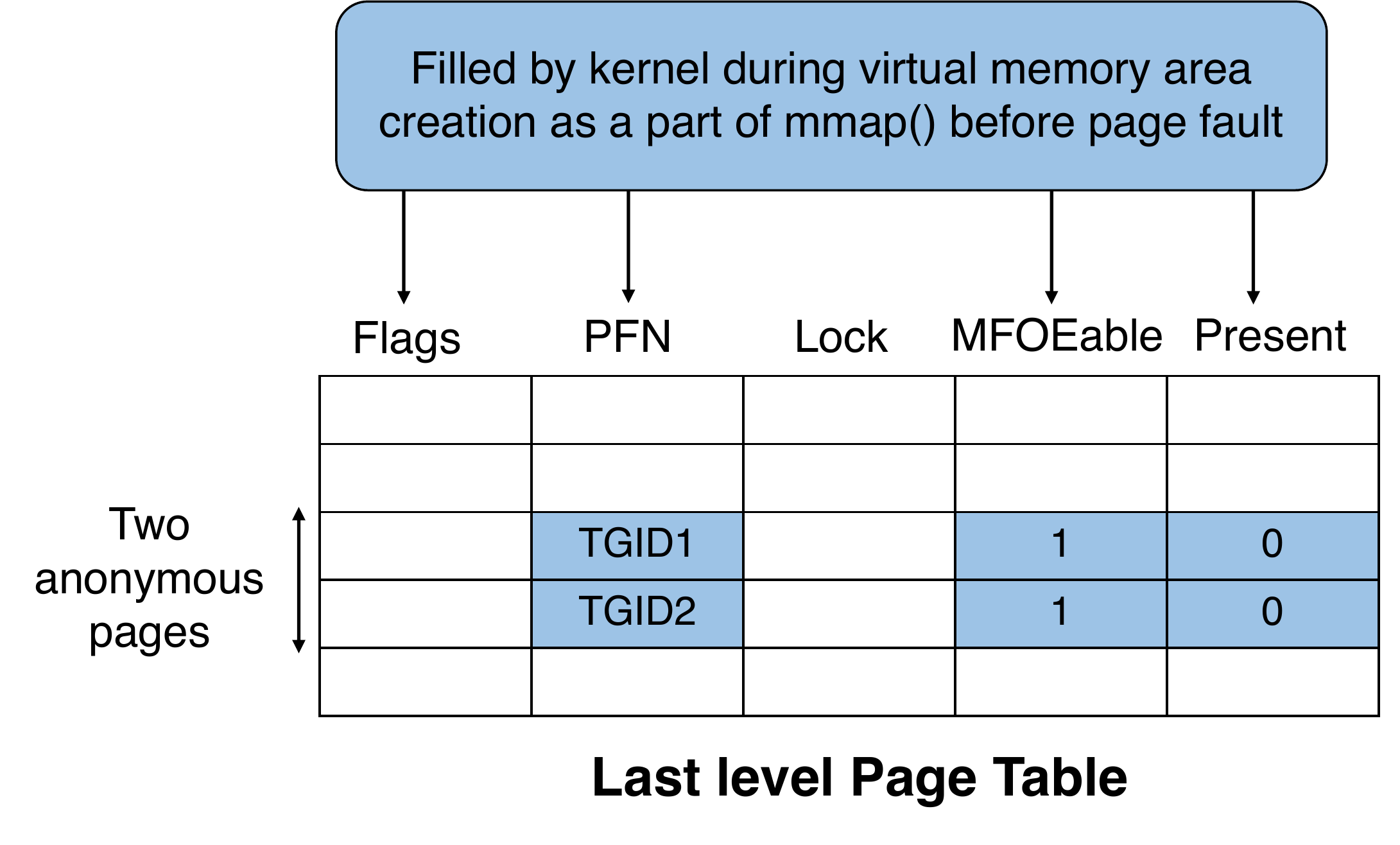}
     	\vspace{-1em}
	\caption{Context of the lowest level PTEs of a MFOE enabled process. The \texttt{mmap} function maps two pages and sets the TGID and MFOEable bits.}
	\label{fig:prepagefault}
	\vspace{-1em}
\end{figure} 
In addition to setting the \texttt{MFOEable} bit, the kernel also fills the RW bit (bit-position 1, if the region is writable),  PFN field of PTE with thread group identifier(TGID) of current process. 
Here we are re-purposing bits of an otherwise empty/invalid PTE to convey the MFOE hardware about the legality of address, the RW permissions, and  the TGID of the process that enabled MFOE. This use of PTE is safe and justified because these bits are not used for swap computation. In a MFOE disabled process, these bits would have been all zeros, hence their use is safe.  Fig.~\ref{fig:prepagefault} shows the status of the lowest level page table entry after the execution of pre-page fault operations.

The page table construction doesn't add overhead to the program performance, as it has to be done sometime through its run-time. If not done now, the table will be constructed as part of the fault handling routine. 
We should also highlight that this page table entry creation could be done by the kernel background thread  spawned through \texttt{kthread\_run()} function. In this implementation we have made a choice of using a background thread to minimize blocking time in user-space. Plus, using a background thread could be advantageous for large programs with high virtual memory footprint. With background thread it is possible that MFOE reaches a PTE of a page which is \texttt{MFOEable} but its \texttt{MFOEable} bit has not been set by the asynchronous kernel thread. In this case, the MFOE will simply treat this page as ``non-MFOEable'' and trigger a typical page fault.

\subsection{Hardware Page Fault  Handling}
\label{subsec:MFOEpagefault}
\begin{figure}[h!]
	\centering
     	\includegraphics[scale=0.375]{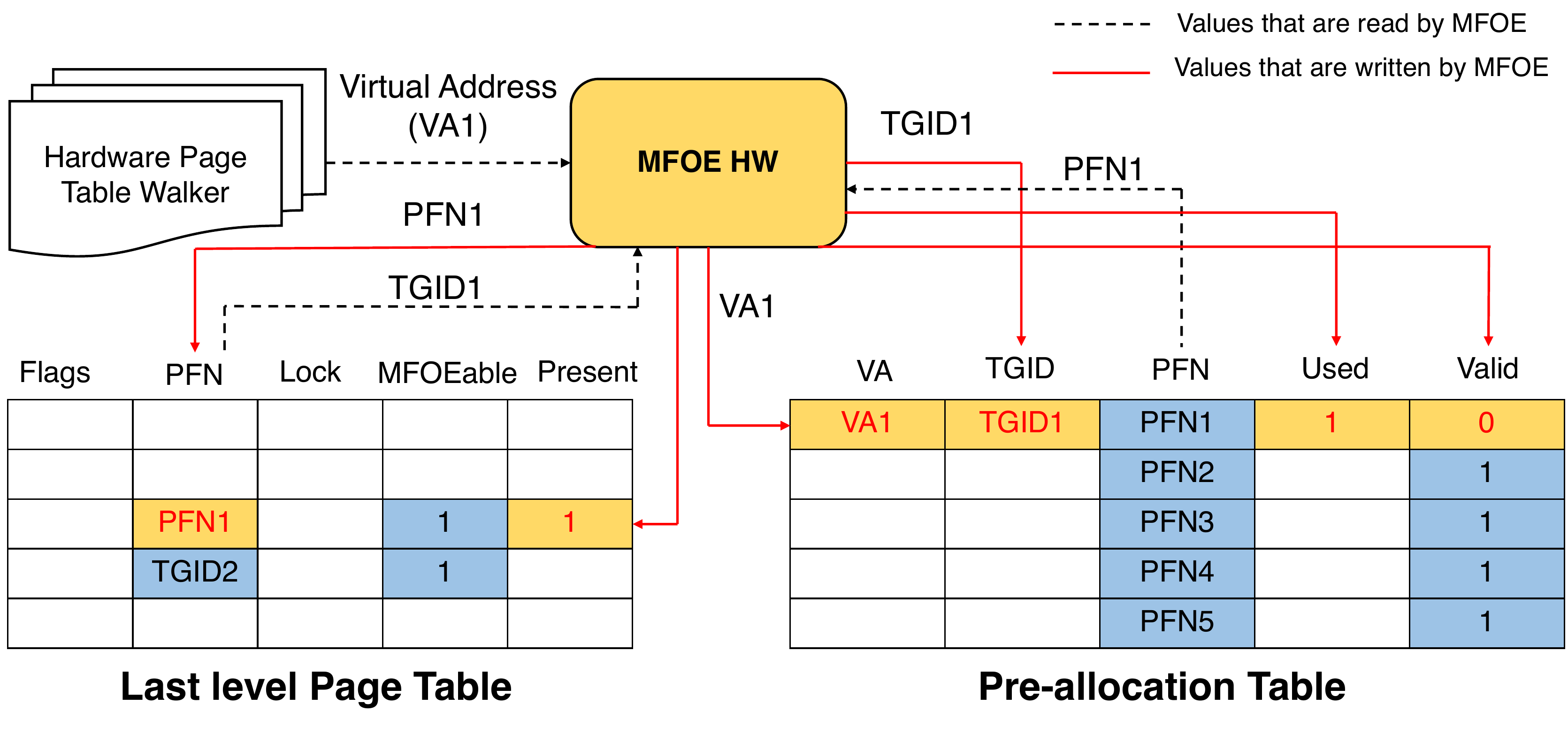}
     	\vspace{-1em}
	\caption{The operations of MFOE. The left table stands for the lowest PTEs and the right table is the pre-allocation table.}
	\label{fig:pagefault}
	\vspace{-1em}
\end{figure}
The MFOE is only activated after a TLB miss and after the hardware page table walker reaches the last-level of PTE. If the hardware page table walker recognizes the present bit is absent in PTE, it activates the MFOE to handle the page fault

The operation of MFOE is as follows:

\begin{enumerate}
\item  If MFOE sees that both the present bit and the \texttt{MFOEable} bit of PTE are not set, this means that this page cannot be pre-allocated or this faulting address might be illegal. MFOE triggers the  page fault exception in this case.
\item  If MFOE sees the present bit of PTE is not set, but the \texttt{MFOEable} bit of PTE is set, this means that this PTE belongs to a legal virtual address. MFOE can now  proceed to next step of handling the minor fault.
\item MFOE then retrieves the starting (physical) address of pre-allocation table by reading the register CR9 from its context. The retrieved address is used to read the header of pre-allocation table. The header read from pre-allocation table reveals the tail index into pre-allocation table.
\item MFOE computes an entry into pre-allocation table, with address to pre-allocation header and tail index retrieved from the header. With this address, MFOE reads an entry of pre-allocation table.
\item The entry read returns a 64-bit word which comprises of valid-bit of the pre-allocated entry. If the valid bit is set, then MFOE can proceed to next step of page fault handling. If not, this means that all the entries in pre-allocation table have been consumed and MFOE has to raise the page fault exception for the fault to be handled by kernel.
\item In the next step, MFOE writes back the pre-allocation entry with updated TGID, VA fields along with used-bit set and valid-bit unset. MFOE used the TGID (written by kernel at section \ref{sec:MFOEprepagefault} 
It also modifies the PFN field of PTE with PFN obtained from pre-allocated entry. MFOE also increments the tail index of the pre-allocation table and writes back the updated tail-index to pre-allocation table header. 
\item MFOE also updates the PTE  here itself, without needing to do the page walk again. It also makes an entry in the TLB to prevent a page walk when the access is retried by the core after fault handling.
\end{enumerate}
Fig.~\ref{fig:pagefault} shows the operations of MFOE during a page fault.


\subsection{Post-Page Fault Software Handling}
\label{subsec:MFOEpostpagefault}

\begin{figure}[h!]
	\centering
     	\includegraphics[scale=0.4]{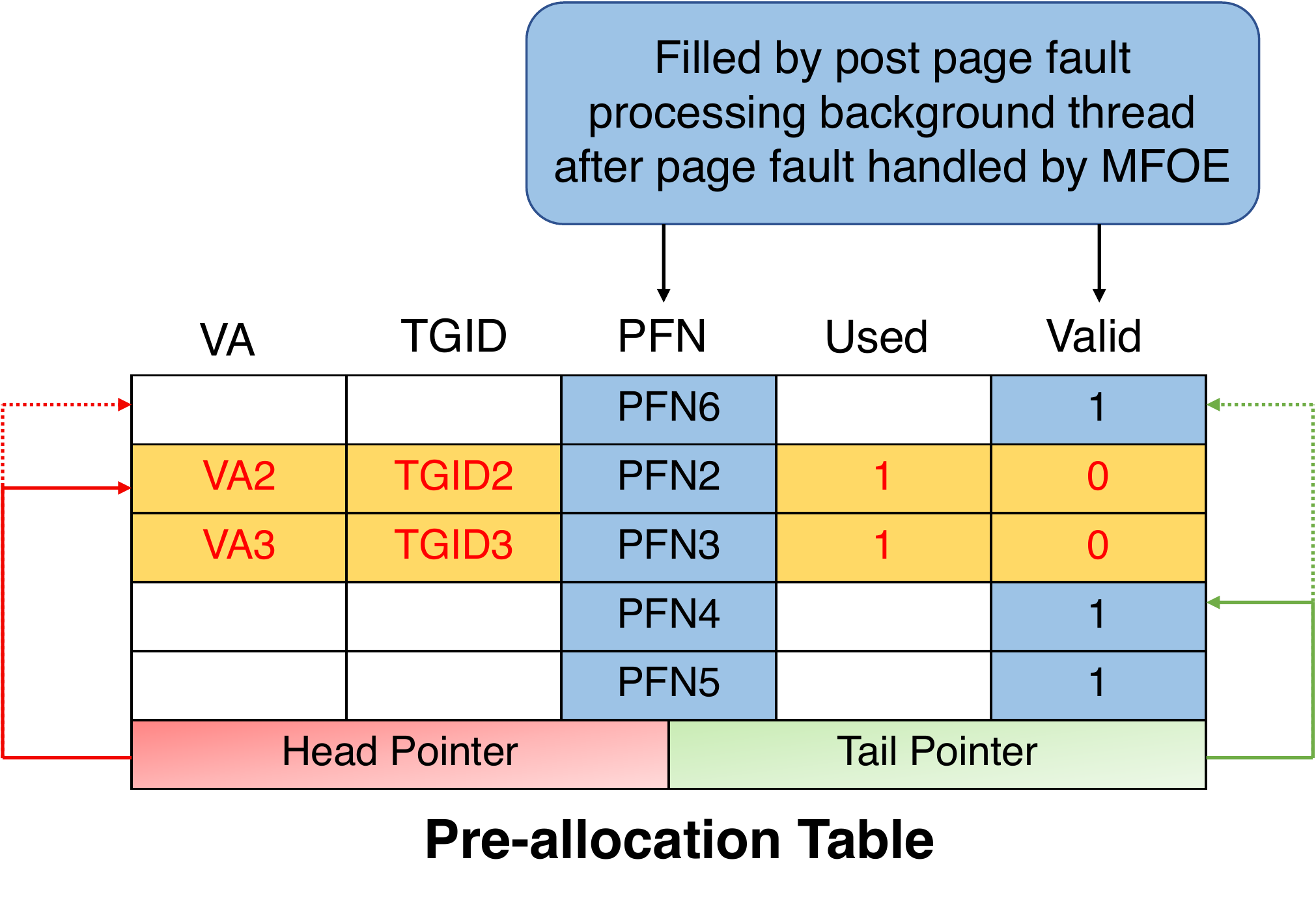}
     	\vspace{-1em}
	\caption{The operations of post-page fault handling.}
	\vspace{-1em}
	\label{fig:postpagefault}
\end{figure}
We used the delayed work queue of the Linux kernel to periodically execute post page-fault processing function. After the pages of all pre-allocation tables are filled by background thread  
(triggered by \texttt{MFOE\_enable} system call), the post page-fault processing function is inserted into the work queue with a programmed delay. This programmed delay dictates how frequently the post page-fault processing should be done. For our implementation we have set it to 2 milli-seconds. 

The following tasks are executed in-order in the post page-fault processing function, for each per-core  pre-allocation table.  
\begin{enumerate}
    \item First, we use the contents of CR9 register to access the pre-allocation table header. Then we use the head field of the header to index into the corresponding head entry of the pre-allocation table. If the head entry's used bit is set, we proceed to the next-step. If the used bit is not set, it means that the MFOE hasn't used the pre-allocation table, possibly indicating that there are no faults since the last pre-allocation table refresh.
    \item From the used pre-allocation table entry, we obtain the TGID, faulting virtual address, and page frame number. Using these values we sequentially call \texttt{anon\_vma\_prepare}, \texttt{inc\_mm\_counter}, \texttt{page\_add\_new\_anon\_rmap},  and \texttt{lru\_cache\_add\_active\_or\_unevictable} functions.  After all the functions are executed, all the fields of the pre-allocation table entry are cleared.
    \item We now allocate a physical page to replenish this head entry. The frame number of the the new page is written to the PFN field, along with the valid and used bits set to 1 and 0 respectively. The head index is incremented by 1 and all the tasks are repeated for the next entry.
\end{enumerate}
Fig.~\ref{fig:postpagefault} illustrates the operations of post-page fault processing function and how it updates the entries of the pre-allocation table.

Since we are performing post page-fault processing once for a programmed delay time (2ms), it is possible that during this interval, the program might have exhausted all the allocated pages. In this case, the MFOE would raise an interrupt to the kernel and fall back to typical page fault handling. Plus, In our current implementation, the post page-fault processing is single threaded, but implementing a multi-threaded version is plausible.

The current implementation of the post page-fault processing does leave room for optimization. For example, instead of periodically triggering the post page-fault processing function, we could dynamically awaken the function based on a pre-allocation pages available for the application. Also based on the usage, one could dynamically increase/decrease the number of the pre-allocation pages for each core. 
As the post page-fault processing function is working to replenish the entries of pre-allocation table of one core, the other core might be starving for pre-allocated pages, resulting in the latter core suffering longer latency in its minor faults. This effect can potentially lead to fairness issues in a  multi-core system. To mitigate this to some extent in the current implementation, we use a round-robbin policy to select the cores for pre-allocation.

\subsection{Error Handling}
\label{subsec:MFOEErrorHandling}
Since the post page-fault processing function executes synchronously once every programmed delay period, there can be a situation where an application is terminated while some of its pages (allocated through MFOE handled page faults) are yet to be processed. This scenario becomes worse, if some of the pre-allocated pages are still unused and needs to be reclaimed by the kernel.

To handle such scenarios, we implement an error handling function that will be called when the application is getting terminated. This function will scan pre-allocation tables of all the cores to find the used entries (with used bit set) whose TGID is the same as the terminated application.  If such entries exist, then the error handling function executes the same set of functions as the post page-fault processing function.  Also, since the error handling function and the post page-fault processing function might execute at the same time, we need a lock bit to avoid race condition between them. The bit 0 of locks field in the pre-allocation table header is employed as a \texttt{test\_and\_set} lock bit here. 

Because the error handling function needs to scan all pre-allocation tables,
it should only scan when necessary. Therefore, the error handling function is only called by the zap\_pte\_range() function and when the page's page\_mapcount() returns zero as well as the vma's VM\_MFOE flag is set. 
It is also worth mentioning that if the parent thread (the thread whose PID is also the TGID) of a thread group is killed before its child threads, then the parent thread must be responsible for handling the unprocessed pages of pre-allocation tables before kernel selects a new thread group leader.
\vspace{-1em}
\subsection{Disabling MFOE}
\label{subsec:MFOEDisable}

The \texttt{MFOE\_disable} function sets the \texttt{mfoe} member of calling process's memory descriptor(\texttt{mm\_struct}) object to false. Like \texttt{MFOE\_enable}, the last caller disables MFOE hardware of each core by calling \texttt{schedule\_on\_each\_cpu()} function to clear the MFOE\_ENABLE bit in CR9 register for all cores, and it also frees/releases all valid pages (whose valid bit is set) from all the pre-allocation tables.

As discussed in section \ref{subsec:MFOEenableanddisable},
MFOE is implemented to be automatically disabled by the \texttt{MFOE\_disable()} function from \_\_mmput()  kernel function, which is called  when the applications are terminated. So disabling MFOE within the source code of applications is unnecessary. Nevertheless we implemented the \texttt{MFOE\_disable} system call to give the applications the flexibility to disable MFOE when needed. Also, we implemented a protection code to ensure that calling \texttt{MFOE\_disable()} function twice accidentally will not harm. When called twice, the second \texttt{MFOE\_disable} system call would directly return.
\vspace{-1em}
\subsection{Software Emulation of MFOE}
\label{subsec:MFOESE}
The critical path of minor page fault handling can also be implemented in software. Hence we implement the software emulation of MFOE in the form of \texttt{mfoe\_se()} system call. \texttt{mfoe\_se} is called with the faulting virtual address as input in the micro-benchmark used in the evaluation. Although, this implementation needs the programmer to have explicit knowledge of faulting address before hand, the functionality of this system call could be easily integrated with typical page fault handling routine(inside \texttt{do\_anonymous\_page())} in future implementations. The software emulation of MFOE is made possible because of the kernel's capability to track both the page tables and pre-allocation tables.

The implementation of \texttt{mfoe\_se()} is as follows
\begin{enumerate}
    \item First we obtain the memory descriptor of the process. Then we hold the non-exclusive access to \texttt{mmap\_sem} using \texttt{down\_read}. We get the virtual memory area descriptor(\texttt{struct vm\_area\_struct}) by calling \texttt{find\_vma()} function with current memory descriptor and faulting address as inputs.
    \item After obtaining the virtual memory descriptor, we check if its flagged with \texttt{VM\_MFOE}. If it is not, then release the \texttt{mmap\_sem} and return with error. If it is flagged as required, we proceed to next steps.
    \item Now we start the multi-level page walk through a series of kernel function calls \texttt{pgd\_offset}, \texttt{pud\_offset},  \texttt{pmd\_offset}. We read the leaf entry in the page table-pte by obtaining the spinlock to the entire last level page table, by using \texttt{pte\_offset\_map\_lock}. At this stage we hold exclusive access to the last level page table.
    \item Now we read the pte to check if it has MFOE'able bit enabled, no valid translation present, and with thread group identifier(TGID) filled in the PFN field. If either of this fields are not present, we release the spinlock and \texttt{mmap\_sem}, return with error. Else we proceed to the next steps.
    \item After checking the validity, We access the pre-allocation table of the faulting core(core number is obtained by \texttt{smp\_processor\_id()} kernel function). We access the entry in the pre-allocation table that's pointed out by the tail pointer. If the entry is available(used bit as 0 and valid bit as 1), then we proceed to next step. Else we release all the locks, return with error.
    \item Now we take the TGID present in the pte, fill it in the tail entry of the pre-allocation table. We also update the pre-allocation table tail entry's virtual address field, used, valid fields. Subsequently, we update the pte with PFN from pre-allocation table. set the used and valid bits to 1 in the pte. 
    \item We release both the page table spinlock and \texttt{mmap\_sem}, return with success to the faulting process.
\end{enumerate}
\vspace{-0.5em}
At this stage, when the virtual address is accessed through load/store it won't generate a page fault, but will result in a TLB miss. The hardware page table walker performs the walking to the get the translation we create through the \texttt{mfoe\_se} system call, caches it in TLB.
\vspace{-1em}
\subsection{Other Issues}
\label{subsec:MFOEOtherIssues}

\textbf{Resource Limitation:} Each process in Linux can only allocate a limited amount of resources, such as CPU time, memory, and files. Although it is rare, it is still possible that the MFOE  enabled process would allocate pages more than its limitation  because MFOE hardware cannot check during a page fault. To prevent this case, the periodic post page-fault background thread needs to check the allocated pages and the quota of pages for this process. If the total allocated pages of a process are over some threshold, say 80\% of main memory,  this background thread will call \texttt{MFOE\_disable()} function for this process and
triggers another background thread to clear the \texttt{MFOEable} bit of PTEs of all unprocessed pages of this process. 

\textbf{Huge Page Support:} MFOE can also support huge pages easily. The current architecture can be re-purposed to use huge(2MB) pages instead of base(4KB) pages easily. If the intention was to use both 4KB and 2MB pages for handling the faults, another pre-allocation table per core is needed to contain all pre-allocated huge pages for a core. The kernel background thread in section \ref{subsec:MFOEprepagefault} will set the \texttt{MFOEable} bit at PMD if huge page can be allocated for certain virtual address regions. When MFOE reaches PMD and finds the MFOEable bit is set but present bit is not set, MFOE will obtain a huge page from the pre-allocated table. 

Our approach in this paper provides some advantages similar to the use of huge(2MB) pages, while managing the memory with base(4KB) pages. For example, using huge pages would reduce the number of faults that the application has to endure, but the kernel would have to deal with promoting/demoting huge pages and also perform memory compaction to make sure that it doesn't run out of huge pages. On the other hand, with MFOE, the application would have to endure more faults, but the latency of each page fault now would be reduced and subsequently the kernel has to do less amount of compaction.


{\bf Virtual Machine Support}

Prior work shows that the memory translation in the virtual machine environment becomes two-dimensional~\cite{DIY,walker} i.e. each physical address in the guest will be a virtual address in the host. TLB in a virtual system caches the guest virtual address to host physical address translations. A TLB miss would trigger the 2-D hardware page table walker, which walks both the guest and host page tables. MFOE in a virtual system environment will be implemented as optimizations to 2D page table walker.

We outline one possible way of implementing MFOE in hardware assisted virtualization. The guest OS will have its software changes in Linux kernel, as if its run on a real machine. The guest OS would construct the page table walking path and set the MFOEable bit in leaf entry of the page table, as explained in \ref{subsec:MFOEprepagefault}. Since the virtual machine is like any process for the host, the host OS(hypervisor) also constructs the page table walking path and set the MFOEable bit for the lowest PTE. Both guest OS and hypervisor would maintain their own pre-allocation tables, with MFOE being able to access both of them.

In an event of TLB miss, the page table walker starts to walk the 2D path till it reaches the lowest PTE in guest page table first. If MFOEable bit is set, MFOE will be consulted to obtain a page from the guest pre-allocation table. But the walk won't stop here, it continues till the walker reaches the lowest page table entry of the host. Again here, if the MFOEable bit is set, MFOE is consulted now to get the physical page number from the host pre-allocation table.  Finally, the MFOE updates the corresponding TLB entry and re-executes the faulting instruction.  We should mention that we should avoid breaking the translation path from pages in the guest pre-allocation tables to its physical address in the hypervisor.  The ballooning driver and the hypervisor swapping mechanisms \cite{vmware} adopted in current hypervisors might have to be disabled for the physical pages in hosts that support pre-allocated pages in guest OSs. We leave this virtual machine support as a future work.


\textbf{Multi-thread support:} 
With multi-threaded applications, two or more threads might access the same page and therefore encounter the page fault (of the same page) at the same time. Furthermore, we should consider a more complicated race condition case between MFOE hardware  and kernel software page fault exception handler. In section \ref{subsec:MFOEprepagefault}, a kernel background thread is used to set the MFOEable bit for all PTEs. Consider the following rare case, the MFOE hardware reaches a PTE before the background thread sets its MFOEable bit. Therefore, MFOE hardware will treat this page as ``non-MFOEable'' and triggers a page fault exception. Before the page fault exception handler executes, our kernel background thread could be scheduled and sets  the MFOEable bit of this PTE.  MFOE hardware from another core could reach this PTE  (since its present bit is not set but MFOEable bit is set) and execute the MFOE page fault handling. 
Meanwhile, the page fault exception handler triggered from the first core executes and starts to handle the page fault for the same faulting page. 
This hardware and software race condition might be a serious problem, so we have to avoid it.

To solve this problem, we employ a bit lock at the lowest PTE.
That is, when the MFOE hardware and kernel page fault exception handler want to service a page fault\footnote{Kernel can check the VM\_MFOE flag from vma to decide if it needs to get the lock or not. If VM\_MFOE is not enabled, this locking step is skipped.}, they both need to obtain a lock at the lowest PTE first.  We re-purpose avl bit of the empty PTE as the lock bit\footnote{This bit is not used in an invalid PTE for a typical system. and is not used for detecting if the page is swapped. Hence, this use is justifiable}. In software,\texttt{test\_and\_set} operation is used by fault handling method to atomically set the lock. If the kernel sees that lock bit is already set, this means that MFOE from some hardware context is handling the fault. It is justifiable for kernel to busy wait on a bit lock, as the MFOE fault handling takes very small amount of time.

The page fault handling flow mentioned in \ref{subsec:MFOEpagefault} becomes as follows. Once the MFOE is able to obtain a free page from pre-allocation table. MFOE must do a \texttt{read\_modify\_write} atomic operation on the PTE to acquire the lock. This is accomplished in hardware using locked memory read and write operations. MFOE does a locked memory read on PTE, resulting in blocking of any other core trying to read the same PTE. If MFOE sees the lock bit is 0, it does a locked write to PTE setting the lock bit set to 1. The other core which have been blocked trying to the read the same PTE will now read the PTE with lock bit as 1. 
If the\texttt{read\_modify\_write} operation is successful, MFOE proceeds to next step of page fault handling. After the page fault has been handled, MFOE will clear the lock bit to unlock the PTE.
However, if initial locked read returns 1 on lock bit, this means that MFOE from another core 
or kernel page fault exception handler is handling a page fault for the same page, 
then MFOE would need to busy wait here until the lock bit is clear and the present bit is set (this means that the page fault has been serviced),  and then updates TLB as well as re-execute the faulting instruction. 

This architecture is more scalable compared to the existing kernel exception handler approach. MFOE only busy waits at the page level, and this busy waiting happens only when multiple MFOEs or the kernel exception handler try to access the same page simultaneously.  However, the existing kernel page fault exception handler uses a global page table lock to  synchronize all page faults of the threads belonging to the same process. Therefore in a typical system, all faulting threads of the same process must busy wait no matter their faulting pages are the same or not.



%% file: eval.tex
\section{Evaluation}
\label{sec:POEevaluation}

Here we first discuss our evaluation methodology, followed by a discussion of results on  micro-benchmarks and  full applications.
\vspace{-1em}
\subsection{Methodology}
\label{subsec:methodology}
We evaluate the full hardware/software MFOE technique against traditional minor fault handling in OS software baseline.  To explore the benefits of the proposed MFOE enhanced hardware page walker, we also examine a software emulation of MFOE, without the enhanced hardware page walker, implemented as a system call.  These systems are evaluated under a combination of large memory footprint  applications as well as a set of micro-benchmarks used to aid in the analysis of the system.  
For full application workloads we select applications taken from several benchmark suites like  GAPBS\cite{beamer2015gap}, Memcached as database layer of YCSB~\cite{10.1145/1807128.1807152}, SPLASH and PARSEC~\cite{10.1145/3053277.3053279}. The reason we have picked only few applications from their respective benchmark suites is that these applications have considerable minor fault overhead for evaluating MFOE. We also wanted the evaluated benchmarks to be diverse. More information about each benchmark and their configuration can be found in Table.\ref{table:benchmark_description}. All the applications are run in multi-threaded mode(if possible) and parallel fault events are taken into consideration, when calculating minor fault overhead.
\begin{table}[h!]
    \centering
    \begin{tabular}{|l|c|}\hline
        \textbf{Benchmark} & \textbf{Description} \\ \hline  
         GCC compiler & Representative of Compiler workloads. \\ & Linux kernel 4.19 compilation using gcc(multi-threaded mode)\\ \hline
         Parsec-dedup & Multi-threaded data de-duplication benchmark taken from \\ & PARSEC\cite{PARSEC} benchmark suite \\ \hline
         Function-as-a-service & Faas-profiler \cite{10.1145/3352460.3358296} generated synthetic traffic-20 invocations/sec \\
         & No pre-warm containers. Container memory limit = 128MB \\ \hline
        YCSB-memcached & Memcached used as database layer for YSCB~\cite{10.1145/1807128.1807152}. \\ & Workload used is 100\% insert. \\ & Memcached server is running on 4 threads \\ \hline
         Splash2X-radix & Multi-threaded radix sort benchmark taken from SPLASH2X\cite{SPLASH2}.\\ & Input size is chosen to be "native"\\ \hline
         Splash2X-fft & Multi-threaded fourier transform benchmark taken from SPLASH2X\cite{SPLASH2}.\\ & Input size is chosen to be "native".\\\hline
         XSBench & Mini-app representing key computational \\ 
         & kernel of the Monte Carlo neutron transport algorithm\cite{Tramm:wy}. \\ & Used input size "large" \\ \hline 
         Integer Sort & Integer sort benchmark taken from NPB\cite{npb}. \\ \hline
         
    \end{tabular}
    \caption{Benchmarks Description}
    \label{table:benchmark_description}
    \vspace{-1em}
\end{table}

Because MFOE proposes both hardware and software changes to the system, we must simulate the full proposed system to accurately model the impact of the proposed changes. The primary benefit of MFOE, however, will be seen on workloads with large memory utilization, which makes simulation challenging.  Thus, to evaluate the performance of our proposed MFOE system, we take a combined approach of detailed microarchitectural simulation together with real system tracing and analytical modeling.

\begin{table}[h]
\begin{subtable}[t]{0.475\textwidth}
\begin{tabular}{|c|c|}
\hline
Architecture & x86\_64 \\ \hline
CPU Model & TimingSimpleCPU \\ \hline
No. of cores & 8  \\ \hline
Cache-levels & 2 \\ \hline
Clock-speed & 3GHz \\ \hline
L1 I-Cache & 32 KB \\ 
& 4-way set-associative\\ 
& access-latency: 2 cycles \\ \hline 
L1 D-Cache & 32 KB \\
& 4-way set-associative \\
&  access-latency: 2 cycles \\ \hline
L2 Cache & 256 KB\\
& 8-way set-associative \\ 
& access-latency: 8 cycles \\ \hline
Cache Coherency & MOESI Hammer \\ \hline
Main Memory  & 16GB DDR4 2400 MT/s\\ \hline
Kernel & Linux 4.9.182 \\ \hline
\end{tabular}
\caption{gem5 simulated system configuration}
\label{table:gem5}
\end{subtable}
\hspace{\fill}
\begin{subtable}[t]{0.475\textwidth}
\flushleft
\begin{tabular}{|c|c|}
\hline
Architecture & x86\_64 \\ \hline
CPU Model & Intel Xeon E-2136 \\ \hline
Clock-speed(max) & 3.3GHz \\ \hline
No. of cores & 6  \\ \hline
No.of Threads & 12 \\ \hline
Cache-levels & 3 \\ \hline
L1 I-Cache & 32 KB \\ 
& 8 way set-associative \\ \hline 
L1 D-Cache & 32 KB \\ & 8 way set-associative \\ \hline
L2 Cache &  256 KB \\ & 4 way set-associative\\  \hline
L3 Cache & 12 MB \\  & 16 way set-associative\\ \hline
Main Memory  & 32 GB DDR4 \\ & 21300 MT/s\\ \hline
Kernel & Linux 4.9.182 \\ \hline
\end{tabular}
\caption{real system configuration used for evaluation}
\label{table:realsystem}
\end{subtable}
\caption{system configuration used for evaluation}
\vspace{-3em}
\end{table}
 For detailed micro-architectural simulation, we implemented MFOE on gem5~\cite{GEM5} in full-system emulation mode as it allows prototyping with both hardware and software.  The simulated gem5 system configuration is as shown in table Table .~\ref{table:gem5}. The gem5 page walker-TLB sub-system was enhanced to implement MFOE.  This system was used to perform microbenchmarking and establish the latency of typical minor page faults as well as MFOE-enhanced minor page faults. The microbenchmark analysis is presented in \ref{subsubsec:latencybenchmarks}

We also implemented a software-only version of MFOE in the form of a system-call(\texttt{mfoe\_se}) whose implementation was explained in \ref{subsec:MFOESE}. We evaluated \texttt{mfoe\_se} on real system whose configuration is detailed in Table.\ref{table:realsystem}

Since gem5  is too slow for memory intensive applications to complete in a tenable amount of time, we estimate the gain for  benchmarks as if they are run on the proposed hardware with a background thread refreshing the pre-allocation tables.  This was done via analysis with performance model of a system equipped with MFOE. The performance model takes the workload traces generated from real system running these benchmarks as an input and gives the estimated performance gain and MFOE hit rate.
\begin{figure}[h!]
    \centering
    \includegraphics[scale=0.275]{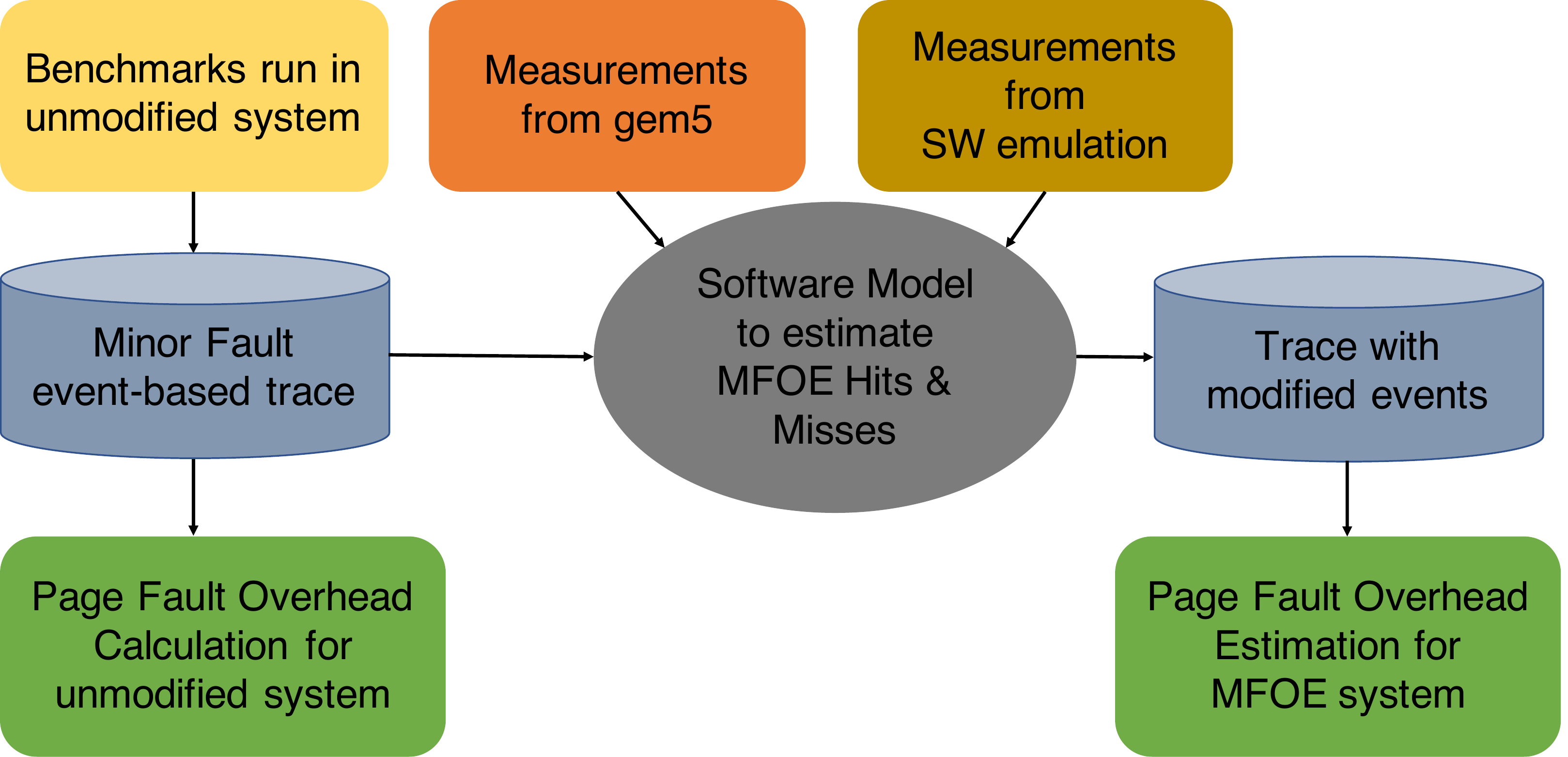}
    \caption{Flow chart of methodology used to estimate benefit due to MFOE}
    \label{fig:estimation_flow_chart}
    \vspace{-1em}
\end{figure}

We incorporated the MFOE hit latency and MFOE miss penalty measurements taken from gem5 simulation in this performance model. Plus, we have measured background thread throughput and initialization throughput from the software only implementation of MFOE. Background thread throughput is the rate at which kernel background  thread will process the MFOE consumed pages and re-allocate frames for future consumption. Initialization throughput is the rate at which the pre-allocation tables are initialized with free pages at the start of application. Initialization throughput is expected to be larger than background thread throughput because the quantum of work for initialization(frame allocation) is smaller than background thread(frame allocation plus kernel meta-data updates). We believe that this performance model will give us best approximation of performance gains due to MFOE.

The working mechanism of our performance model is event based, the event being minor page fault. Each entry in trace we captured from the real system has the core number on which fault has occurred, and the latency of the page fault. The performance model keeps track of pre-allocation tables and their refresh timings have been known due to measurements we incorporated from microbenchmarks. So, given the occurrence of a fault on a core, we check the pre-allocation table for free physical page availability. If its available, we declare the fault as MFOE hit and all the future faults that occur on the faulting core will be moved back in time \texttt{(page fault latency-MFOE hit time)}. If the fault is MFOE miss, We move all the future faults occurring on the same core into the future by \texttt{MFOE miss penalty}. Also, to simulate the producer-consumer nature of software and hardware, we check for faults between each scheduled pre-allocation table update.

Through the explained mechanism, the performance model generates a new trace(timeline) of minor fault events with modified latency. Comparing the fault overheads in original trace, we can estimate the performance gain through MFOE. Overall performance model methodology is shown in Fig.\ref{fig:estimation_flow_chart}. The values of parameters chosen for modeling are shown in Table~\ref{table:Model Parameters}. 
\subsection{Micro-benchmarks}
\label{subsec:microbenchmarks}
\subsubsection{Latency benchmarks}
\label{subsubsec:latencybenchmarks}
We first examine the critical-path latency of MFOE handled anonymous faults, which is the in-line time taken by MFOE to service the page fault.  We call this latency MFOE hit latency.
Not all minor faults which occur when MFOE is enabled will result in a hit. MFOE misses occur when the given thread uses up all the pre-allocated pages before the background post page-fault processing thread replenishes them.  In this case, fault handling reverts back to default Linux kernel.  Because MFOE is enabled, however, there is some wasted time as MFOE first attempts to service the fault. We call this extra time the MFOE miss penalty.

We devised a micro benchmark that creates and strides through a memory region. Between each stride access, the benchmark does a programmable amount of redundant data accesses(non faulting), to simulate the behavior of real program. We do the redundant accesses to simulate different levels of cache availability for the MFOE memory accesses. We made this program multi-threaded i.e. each core will perform faulting accesses and redundant accesses. Each thread generates a total of 32K faults. The fault latency is measured as difference between the clock ticks of start and end events in hardware. The start event being the MFOE invocation by the hardware page table walker, when the walker sees there is no translation for a MFOE'able page. The end event can be either of two possibilities, one possibility is when MFOE indicates to the faulting core to retry the faulting instruction(MFOE hit) and the other being MFOE reverting back to fault handling in software due to lack of pages in pre-allocation table (MFOE miss).


We observe that the  MFOE hit latency on average is 36 CPU cycles, with standard-deviation of 42 CPU cycles. Such large standard-deviation is expected as handling page fault with MFOE involves multiple memory accesses which might or might not hit in CPU caches. The tail latency(95th percentile) of MFOE hit is observed to be 125 CPU cycles. Since the standard-deviation is higher than the mean, we take  78 cycles(mean+SD) as a tighter estimate for MFOE hit latency.
On the other hand, MFOE miss penalty has an average of 14 CPU cycles, with standard-deviation of 5 cycles. The tail latency of MFOE miss is found to be 14 cycles. MFOE miss penalty has little standard deviation because it performs less number of memory accesses compared to MFOE Hit. 

We ran the same program on the real system and analyze the latency of minor page fault handling in unmodified Linux kernel. We measured the wall clock time spent in the kernel function \texttt{do\_anonymous\_page()} which implements the core functionality of anonymous fault handling. The measurement is done using  ftrace utility of the linux kernel. ftrace captures entry and exit instants of \texttt{do\_anonymous\_page} function. The average page fault handling latency in unmodified linux kernel was observed to be 1.27 $\mu$s(2552 CPU cycles), the tail latency is found to be 3.20 $\mu$s(6432 CPU cycles).
Comparing these measurements we find that MFOE improves the average minor page fault handling latency by a factor of 33$\times$ compared to the current systems. The tail latency of minor page fault is improved by 51$\times$ compared to the baseline. Given that the default Linux minor page fault handling latency on average takes around 3 thousand cycles, miss penalty incurred by MFOE is negligible.

It might seem that we are estimating the gain in critical path latency by comparing two different setups. our evaluation choice is justified due to following reasons. Baseline measurements have to be done on real system as the simulation system doesn't reflect the configuration of the processors to-date. While it would have been possible to measure the typical fault handling latency on the simulated system, based on our experience the software timing measurements are more accurate in real systems than in full system simulation. The hardware timing measurement on the other hand are accurate in simulations, so we used those to measure how much time the MFOE takes for handling the page fault. 

\subsubsection{Software-only Implementation of MFOE}
For measuring the latency, we ran the same benchmark as in \ref{subsubsec:latencybenchmarks}, but with \texttt{mfoe\_se()} system call before every faulting access. We measured the time it takes for the \texttt{mfoe\_se} to execute by measuring the entry \& exit points of the system call through \texttt{ktime\_get\_ns()}. Our experiments indicate that
minor faults handled through software emulation of MFOE have an average latency of  795 ns, the tail latency was observed to be 1757 ns. Comparing with the baseline, software emulated MFOE can speed up average fault handling latency by 1.59$\times$ and improves the minor fault handling tail latency by a factor of 1.82$\times$.

These measurements confirm two things for us, (1) Pre-allocation \& lazy kernel data-strcture updates is indeed helpful in reducing the critical page fault handling latency. Even when the simplified critical-path is implemented in software(\texttt{mfoe\_se()}), we are able to achieve latency improvement (2) Hardware/software MFOE approach is superior in reducing minor fault handling critical-path time. Hardware implemented MFOE resulted in 33$\times$ reduction in critical-path latency compared to 1.59$\times$ improvement provided by software only implementation.

We should highlight the performance contrast between hardware offload(MFOE) and software emulation. MFOE poses substantial improvement in fault handling latency because it can handle the page fault in less than 5 memory accesses, which translates to very few cycles depending on cache hits/misses. All the memory accesses that MFOE makes are with physical addresses, so memory virtualization overhead is removed. 
On the other hand, software emulation implementation has to call multiple kernel functions for fault handling, which on aggregate, results in many memory reads/writes. Also, with software emulated implementation of MFOE, there will be a second TLB miss when the access is retried which will result in a redundant page table walk. But with MFOE, there will be a TLB hit when the access is retried. It is noted that software only implementation provides substantial 59\% improvement over the baseline Linux kernel, even though it cannot match the gains from a hardware implementation of MFOE.

\begin{table}[h!]
\begin{center}
\small
\begin{tabular}{|c|c|}
\hline
MFOE Hit Latency & 78 cycles \\ \hline
MFOE Miss Penalty & 14 cycles \\ \hline
Background Thread Throughput & 580.169 KPages/Sec  \\ \hline
Pre-Allocation Initialization Throughput & 1093.075 KPages/Sec \\ \hline
\end{tabular}
\caption{Parameters used for Modelling}
\label{table:Model Parameters}
\end{center}
\end{table}

\begin{figure}[h]
    \vspace{-2em}
     \centering
     \begin{subfigure}[b]{0.49\textwidth}
         \centering
        \includegraphics[scale=0.4]{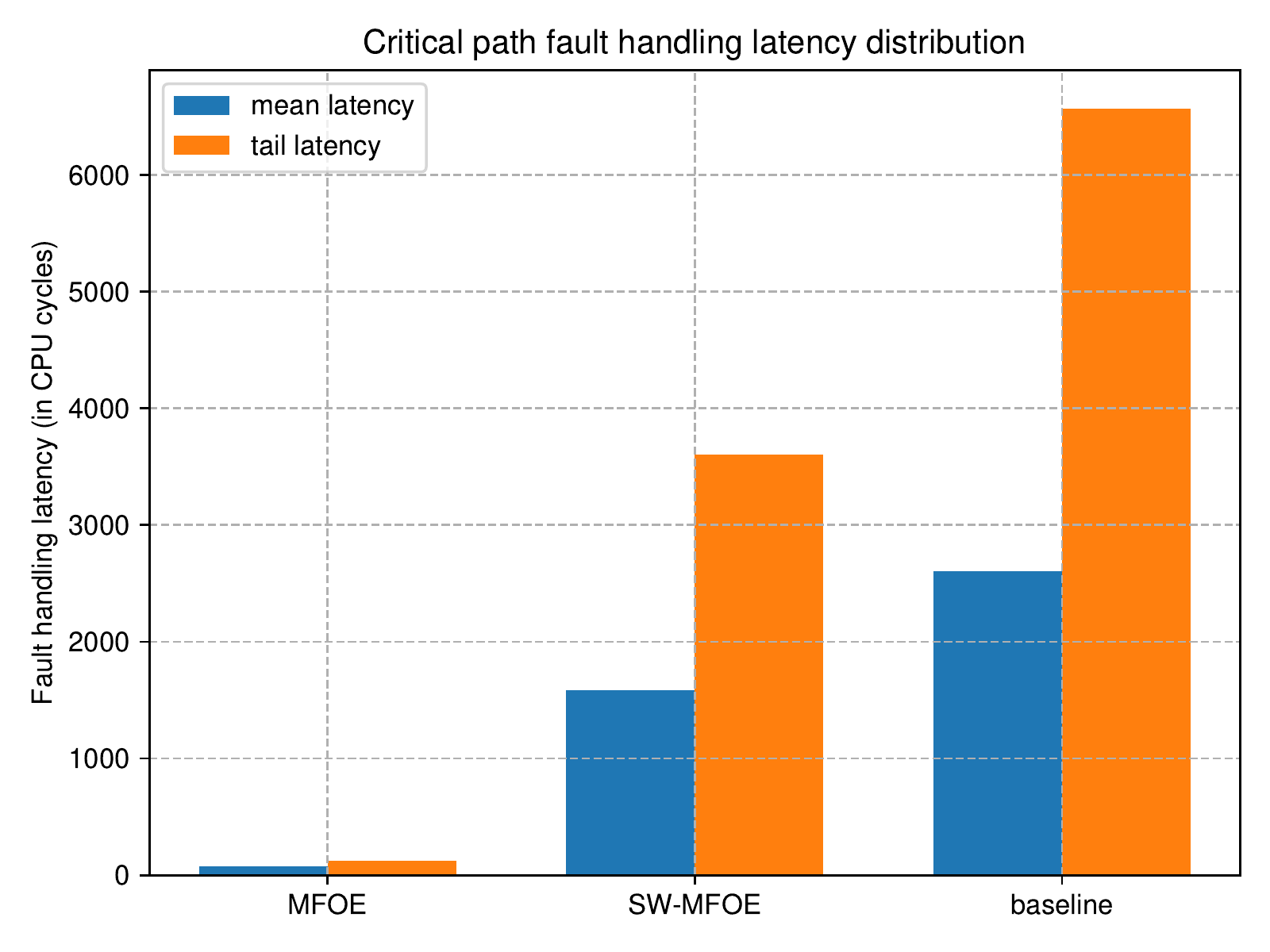}
        \caption{Fault handling of latency MFOE, SW-MFOE, \& baseline}
        \label{fig:mfoe_latency}
     \end{subfigure}
     \hfill
     \begin{subfigure}[b]{0.49\textwidth}
         \centering
         \includegraphics[scale=0.4]{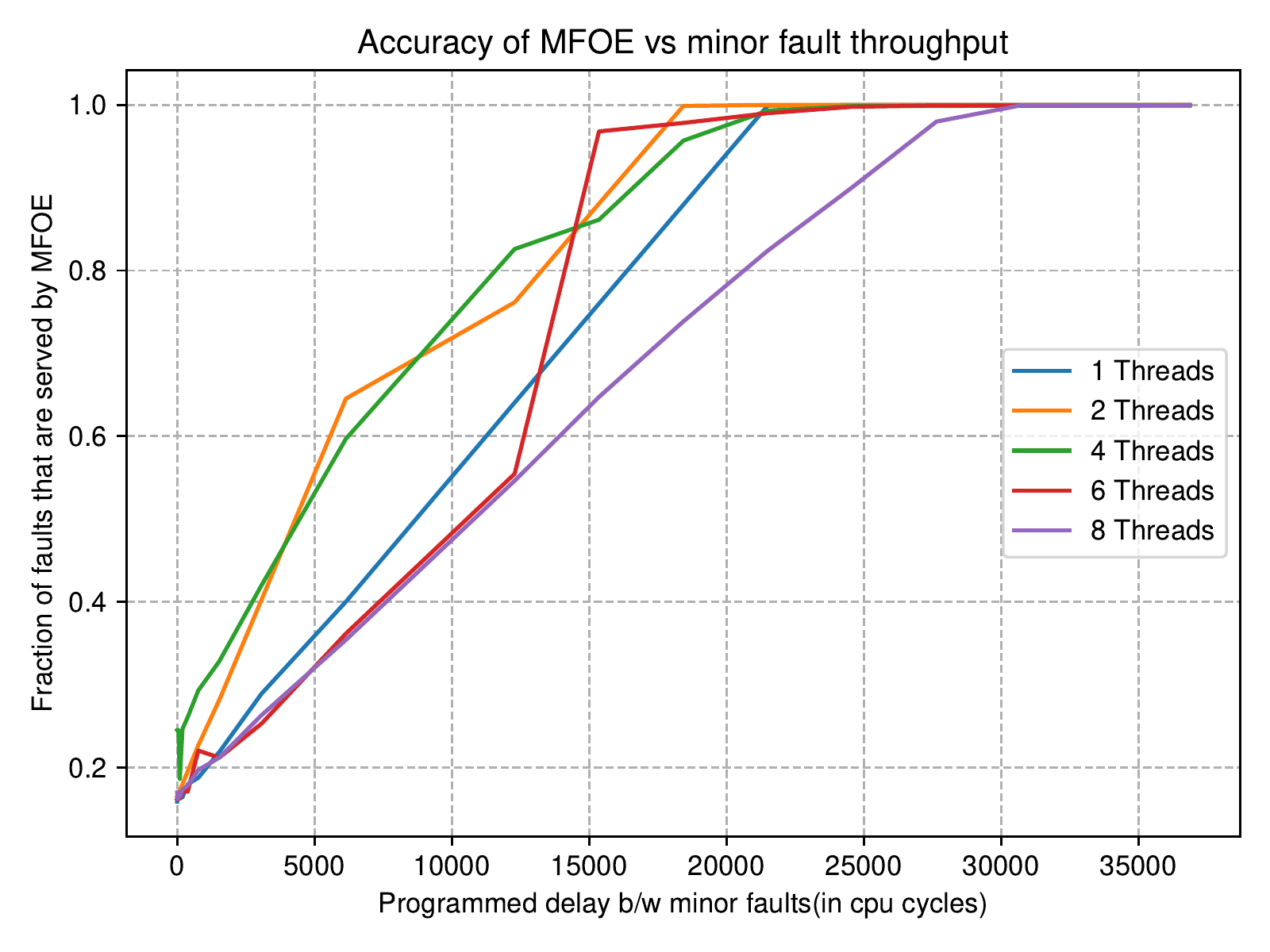}
         \caption{MFOE hit rate for different minor fault occurrence rate}
         \label{fig:mfoe_throughput}
     \end{subfigure}
     \hfill
        \caption{Microbenchmark results}
        \label{fig:microbenchmarkresults}
\end{figure}

Additionally we have devised micro-benchmarks to measure background thread throughput and initialization thread throughput for the software only version of MFOE. These throughput measurements give us an approximation of how fast the pre-allocation tables on this system could be updated, if this system were to implement MFOE. These measurements are shown in Table \ref{table:Model Parameters} and are used in our performance model to estimate the gains due to MFOE.
The comparison of mean \& tail latency of MFOE, SW-MFOE with that of unmodified Linux kernel(baseline) is shown in Fig.\ref{fig:mfoe_latency}. \vspace{-1em}

\subsubsection{Throughput benchmarks}

We also ran a multi-threaded micro benchmark to stress MFOE of given number of CPUs by generating minor faults at an identical rate in parallel fashion.  Here we examine what fraction of minor page faults can be handled under different rates of minor fault occurrence. We perform this experiment on our simulated gem5 system. The results of this experiment are shown in Figure~\ref{fig:mfoe_throughput}. In the figure we observe that MFOE reaches a hit rate of 90\% at a very high maximum page faulting rate of 1 page fault every 21000 cycles. We also observe that the 8-thread workload has lower MFOE Hit rate compared to lower multi-thread workloads in this experiment, this means that the evaluated MFOE configuration (pre-allocation table of 256 pages and post page-fault processing thread sleep duration of 2ms) is unable to keep up with increased page fault rate of 8 threads. Should a given workload need to service page faults at a higher rate, it is straightforward to dynamically schedule this thread more often based on observed thresholds in the pre-allocation table or to deploy more background threads that process pre-allocation tables. Increasing the size of pre-allocation table might also help in tackling with higher page fault rate.

\begin{table}[h!]
\begin{center}
\small
\begin{tabular}{|l|c|c|c|}
\hline
\textbf{App.} & \textbf{Hit rate(\%)} & \textbf{Fault Overhead} &\textbf{Estimated}\\
& & \textbf{ with MFOE(\%)} & \textbf{ Speedup} \\
\hline
GCC compiler & 68.61 & 12.09 & 1.254 \\ \hline
Function-as-a-service & 56.99 & 4.06 & 1.057 \\ \hline
YCSB-memcached & 97.25 & 0.66 & 1.062\\ \hline
Parsec-dedup & 64.10 & 3.87 & 1.049 \\ \hline
Splash2X-radix & 72.92 & 7.46 &  1.109\\ \hline
Splash2X-fft &  42.19 & 7.13 & 1.035\\ \hline
XSBench & 22.91 & 3.95 & 1.01\\ \hline
GAP-BC & 38.15 & 2.41 & 1.010   \\ \hline
Integer Sort & 56.18 & 1.80 & 1.029 \\ \hline
Geo Mean & & & 1.066 \\ \hline
\end{tabular}
\caption{Estimated Performance Improvement}
\label{table:EstimatedImprovements}
\vspace{-3em}
\end{center}
\end{table}

\begin{figure}[h!]
    \centering
    \begin{subfigure}{1\textwidth}
    \centering
    \includegraphics[scale=0.5]{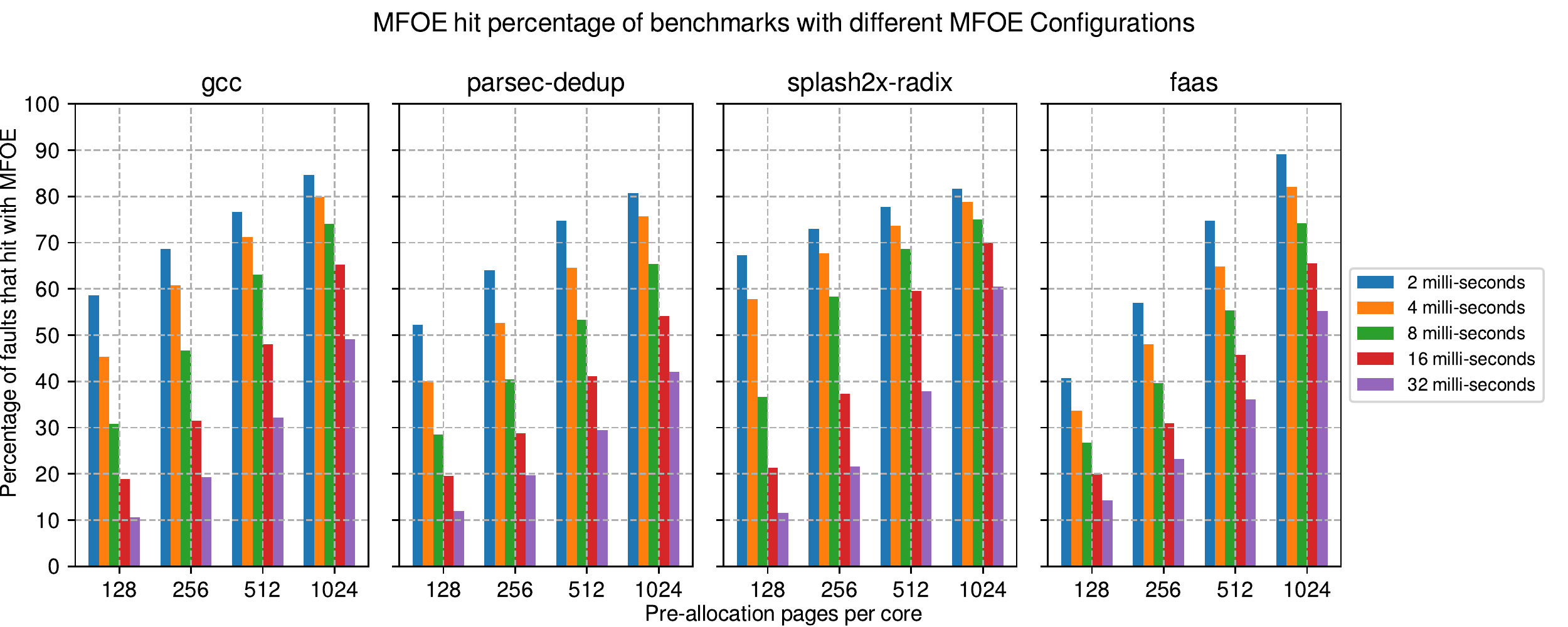}
    \caption{Sensitivity of hit rate with MFOE configuration}
    \label{fig:benchmark_hitrates}
    \end{subfigure}
    \hfill
    \begin{subfigure}[h!]{1\textwidth}
    \centering
    \includegraphics[scale=0.5]{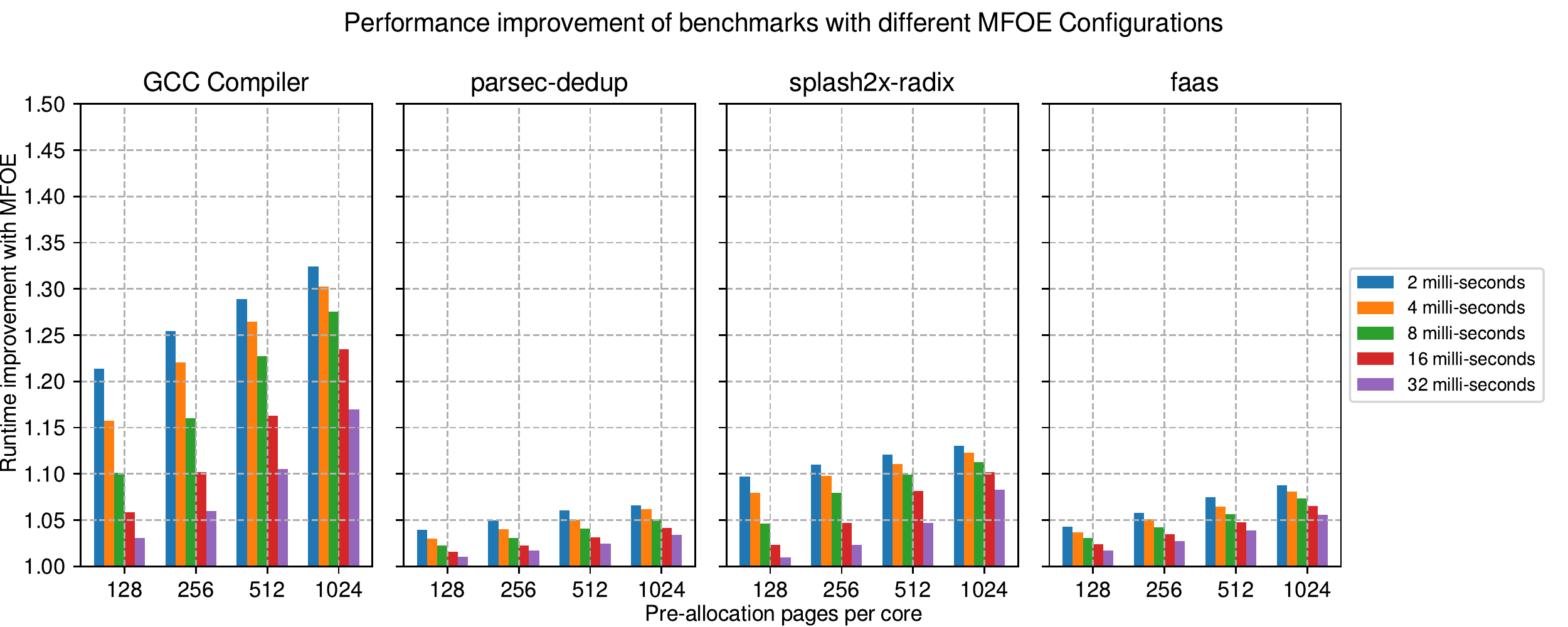}
    \caption{Sensitivity of performance gain with MFOE configuration}
    \label{fig:benchmarks_performancegain}
    \end{subfigure}
    \vspace{-1em}
    \caption{Performance metrics for different benchmarks across different MFOE configurations}
    \vspace{-2em}
\end{figure}
\subsection{Memory Intensive Full Applications}
\label{sec:EstimatedGain}
Table. \ref{table:EstimatedImprovements} shows the MFOE hit rate (i.e. the likelihood that a given minor page fault would be covered by MFOE) as well as the Overhead of minor page faults remaining after MFOE (due to MFOE misses as well as other MFOE overheads) and the total performance gain provided by MFOE on these applications. As the table shows, MFOE provides an average gain of 6.6\% across all the workloads examined. 

Of particular note,\emph{memcached} has a high hit rate of 97\%, thus achieveing run-time improvement of 6.2\% nearly completely removing the overhead due to minor faults. \emph{gcc} has a high 68\% hit rate and thus achieves a nearly 25\% performance gain, effectively capturing most of ideal run-time improvement of 1.40.  By contrast, \emph{splash2x-fft} is only able to achieve a 42\% MFOE hit rate and thus its overheads are higher and only a 3.7\% gain is realized out of a possible 11\%. The same is the case with XSBench which is able to a gain out of 1


To characterize the effect of number of pages allocated during pre-allocation and the background thread refresh-interval, we calculate the performance improvement and MFOE hit rate for 4 benchmarks across different MFOE configurations. The results are shown in Fig.\ref{fig:benchmarks_performancegain} and Fig.\ref{fig:benchmark_hitrates} respectively.
As expected, these results show that MFOE hit rate, and performance gain due to MFOE, decreases as refresh interval of background thread increases. Also, the MFOE hit rate increases with increase in per-core pre-allocation width. The benefits from bigger pre-allocation width and/or faster background refresh, depend on the nature of the application. When the pre-allocation width changes from 512 to 1024, with background thread refresh interval of 2 milli-seconds, faas application had better MFOE hit-rate increment compared to gcc, dedup, and radix applications. This implies that faas application benefits from bigger pre-allocation widths, a possible indication of occurrence of burst faults. On the other hand, for 2 milli-second background refresh, splash2x-radix showed only 21\% improvement in hit rate when pre-allocation width changed from 128 to 1024. This indicates that, radix benefits pre-dominantly from a faster background refresh.
Additionally, under low MFOE Hit rates(for some of the  refresh interval = 32 milli seconds configurations) the applications show lower performance improvement but no performance loss implying that MFOE miss penalty is not substantial compared to performance gains due to MFOE.


%% file: conclusion.tex
\section{Conclusion}
\label{sec:POEconclusion}

Application virtual memory footprints are growing rapidly and to address this growing demand, system integrators are incorporating ever larger amounts of main memory, warranting rethinking of memory management.  These trends are leading to growing overheads due to minor page faults in large memory workloads.  In this paper we proposed the Minor Fault Offload Engine(MFOE) to mitigate this problem through a hardware, software co-design approach.  MFOE parallelizes portions of the kernel page allocation to run ahead of fault time in a separate thread.  This is combined with a hardware accelerator for minor fault handling. MFOE moves much of page fault handling work to background kernel threads and carries out inline work in hardware at the time of the fault.
We show that MFOE can nearly eliminate minor page fault handling costs for most applications and provides performance improvements of up to 25.4\%.

\section{Acknowledgement}
This work was funded in part by an NSF IUCRC grant I/UCRC-1439722 and a grant from Hewlett Packard Enterprise.

%% file: main.bbl

\begin{thebibliography}{37}


\ifx \showCODEN    \undefined \def \showCODEN     #1{\unskip}     \fi
\ifx \showDOI      \undefined \def \showDOI       #1{#1}\fi
\ifx \showISBNx    \undefined \def \showISBNx     #1{\unskip}     \fi
\ifx \showISBNxiii \undefined \def \showISBNxiii  #1{\unskip}     \fi
\ifx \showISSN     \undefined \def \showISSN      #1{\unskip}     \fi
\ifx \showLCCN     \undefined \def \showLCCN      #1{\unskip}     \fi
\ifx \shownote     \undefined \def \shownote      #1{#1}          \fi
\ifx \showarticletitle \undefined \def \showarticletitle #1{#1}   \fi
\ifx \showURL      \undefined \def \showURL       {\relax}        \fi
\providecommand\bibfield[2]{#2}
\providecommand\bibinfo[2]{#2}
\providecommand\natexlab[1]{#1}
\providecommand\showeprint[2][]{arXiv:#2}

\bibitem[\protect\citeauthoryear{??}{bud}{[n.\,d.]}]%
        {buddyallocator}
 \bibinfo{year}{[n.\,d.]}\natexlab{}.
\newblock \bibinfo{booktitle}{\emph{Chapter 6 Physical Page Allocation}}.
\newblock
\urldef\tempurl%
\url{https://www.kernel.org/doc/gorman/html/understand/understand009.html}
\showURL{%
\tempurl}


\bibitem[\protect\citeauthoryear{??}{rin}{[n.\,d.]}]%
        {ringbufer}
 \bibinfo{year}{[n.\,d.]}\natexlab{}.
\newblock \bibinfo{booktitle}{\emph{Lockless Ring Buffer Design}}.
\newblock
\newblock
\shownote{\url{https://www.kernel.org/doc/Documentation/trace/ring-buffer-design.txt}}.


\bibitem[\protect\citeauthoryear{Achermann, Panwar, Bhattacharjee, Roscoe, and
  Gandhi}{Achermann et~al\mbox{.}}{2020}]%
        {10.1145/3373376.3378468}
\bibfield{author}{\bibinfo{person}{Reto Achermann}, \bibinfo{person}{Ashish
  Panwar}, \bibinfo{person}{Abhishek Bhattacharjee}, \bibinfo{person}{Timothy
  Roscoe}, {and} \bibinfo{person}{Jayneel Gandhi}.}
  \bibinfo{year}{2020}\natexlab{}.
\newblock \showarticletitle{Mitosis: Transparently Self-Replicating Page-Tables
  for Large-Memory Machines}. \bibinfo{publisher}{Association for Computing
  Machinery}, \bibinfo{address}{New York, NY, USA}.
\newblock
\showISBNx{9781450371025}
\urldef\tempurl%
\url{https://doi.org/10.1145/3373376.3378468}
\showDOI{\tempurl}


\bibitem[\protect\citeauthoryear{Ahn, Jin, and Huh}{Ahn et~al\mbox{.}}{2012}]%
        {PW-mod1}
\bibfield{author}{\bibinfo{person}{J. Ahn}, \bibinfo{person}{S. Jin}, {and}
  \bibinfo{person}{J. Huh}.} \bibinfo{year}{2012}\natexlab{}.
\newblock \showarticletitle{Revisiting Hardware-assisted Page Walks for
  Virtualized Systems}.
\newblock \bibinfo{journal}{\emph{ACM ISCA Conf.}} (\bibinfo{year}{2012}).
\newblock


\bibitem[\protect\citeauthoryear{Alam, Zhang, Erez, and Etsion}{Alam
  et~al\mbox{.}}{2017}]%
        {DIY}
\bibfield{author}{\bibinfo{person}{Hanna Alam}, \bibinfo{person}{Tianhao
  Zhang}, \bibinfo{person}{Mattan Erez}, {and} \bibinfo{person}{Yoav Etsion}.}
  \bibinfo{year}{2017}\natexlab{}.
\newblock \showarticletitle{Do-it-yourself virtual memory translation}. In
  \bibinfo{booktitle}{\emph{ISCA '17}}. \bibinfo{publisher}{ACM},
  \bibinfo{address}{Toronto, ON, Canada}.
\newblock


\bibitem[\protect\citeauthoryear{Amit}{Amit}{2017}]%
        {203151}
\bibfield{author}{\bibinfo{person}{Nadav Amit}.}
  \bibinfo{year}{2017}\natexlab{}.
\newblock \showarticletitle{Optimizing the {TLB} Shootdown Algorithm with Page
  Access Tracking}. In \bibinfo{booktitle}{\emph{2017 {USENIX} Annual Technical
  Conference ({USENIX} {ATC} 17)}}. \bibinfo{publisher}{{USENIX} Association},
  \bibinfo{address}{Santa Clara, CA}, \bibinfo{pages}{27--39}.
\newblock
\showISBNx{978-1-931971-38-6}
\urldef\tempurl%
\url{https://www.usenix.org/conference/atc17/technical-sessions/presentation/amit}
\showURL{%
\tempurl}


\bibitem[\protect\citeauthoryear{Beamer, Asanović, and Patterson}{Beamer
  et~al\mbox{.}}{2015}]%
        {beamer2015gap}
\bibfield{author}{\bibinfo{person}{Scott Beamer}, \bibinfo{person}{Krste
  Asanović}, {and} \bibinfo{person}{David Patterson}.}
  \bibinfo{year}{2015}\natexlab{}.
\newblock \bibinfo{title}{The GAP Benchmark Suite}.
\newblock
\newblock
\showeprint[arxiv]{1508.03619}~[cs.DC]


\bibitem[\protect\citeauthoryear{Bhargava, Serebrin, Spadini, and
  Manne}{Bhargava et~al\mbox{.}}{2008}]%
        {walker}
\bibfield{author}{\bibinfo{person}{Ravi Bhargava}, \bibinfo{person}{Benjamin
  Serebrin}, \bibinfo{person}{Francesco Spadini}, {and}
  \bibinfo{person}{Srilatha Manne}.} \bibinfo{year}{2008}\natexlab{}.
\newblock \showarticletitle{Accelerating two-dimensional page walks for
  virtualized systems}. In \bibinfo{booktitle}{\emph{International conference
  on Architectural support for programming languages and operating systems
  '13}}.
\newblock


\bibitem[\protect\citeauthoryear{Bhattacharjee}{Bhattacharjee}{2017}]%
        {Translation-prefetching}
\bibfield{author}{\bibinfo{person}{Abhishek Bhattacharjee}.}
  \bibinfo{year}{2017}\natexlab{}.
\newblock \showarticletitle{Translation-Triggered Prefetching}.
\newblock \bibinfo{journal}{\emph{ACM ASPLOS}} (\bibinfo{year}{2017}).
\newblock


\bibitem[\protect\citeauthoryear{Bienia, Kumar, Singh, and Li}{Bienia
  et~al\mbox{.}}{2008}]%
        {PARSEC}
\bibfield{author}{\bibinfo{person}{Christian Bienia}, \bibinfo{person}{Sanjeev
  Kumar}, \bibinfo{person}{Jaswinder~Pal Singh}, {and} \bibinfo{person}{Kai
  Li}.} \bibinfo{year}{2008}\natexlab{}.
\newblock \showarticletitle{The {PARSEC} benchmark suite: characterization and
  architectural implications}. In \bibinfo{booktitle}{\emph{PACT '08}}.
  \bibinfo{publisher}{ACM}, \bibinfo{address}{Toronto, Ontario, Canada},
  \bibinfo{pages}{72--81}.
\newblock


\bibitem[\protect\citeauthoryear{Binkert, Beckmann, Black, Reinhardt, Saidi,
  Basu, Hestness, Hower, Krishna, Sardashti, Sen, Sewell, Shoaib, Vaish, Hill,
  and Wood}{Binkert et~al\mbox{.}}{2011}]%
        {GEM5}
\bibfield{author}{\bibinfo{person}{Nathan Binkert}, \bibinfo{person}{Bradford
  Beckmann}, \bibinfo{person}{Gabriel Black}, \bibinfo{person}{Steven~K.
  Reinhardt}, \bibinfo{person}{Ali Saidi}, \bibinfo{person}{Arkaprava Basu},
  \bibinfo{person}{Joel Hestness}, \bibinfo{person}{Derek~R. Hower},
  \bibinfo{person}{Tushar Krishna}, \bibinfo{person}{Somayeh Sardashti},
  \bibinfo{person}{Rathijit Sen}, \bibinfo{person}{Korey Sewell},
  \bibinfo{person}{Muhammad Shoaib}, \bibinfo{person}{Nilay Vaish},
  \bibinfo{person}{Mark~D. Hill}, {and} \bibinfo{person}{David~A. Wood}.}
  \bibinfo{year}{2011}\natexlab{}.
\newblock \showarticletitle{The gem5 simulator}.
\newblock \bibinfo{journal}{\emph{ACM SIGARCH Computer Architecture News}}
  \bibinfo{volume}{39} (\bibinfo{date}{May} \bibinfo{year}{2011}),
  \bibinfo{pages}{1--7}.
\newblock
Issue 2.


\bibitem[\protect\citeauthoryear{C-C.Chou, Jung, Reddy, Gratz, and
  Voight}{C-C.Chou et~al\mbox{.}}{2019}]%
        {vnvml}
\bibfield{author}{\bibinfo{person}{C-C.Chou}, \bibinfo{person}{J. Jung},
  \bibinfo{person}{N. Reddy}, \bibinfo{person}{P. Gratz}, {and}
  \bibinfo{person}{D. Voight}.} \bibinfo{year}{2019}\natexlab{}.
\newblock \showarticletitle{vNVML: An Efficient Shared Library for Virtualizing
  and Sharing Non-volatile Memories}.
\newblock \bibinfo{journal}{\emph{IEEE Mass Storage Symposium}}
  (\bibinfo{year}{2019}).
\newblock


\bibitem[\protect\citeauthoryear{Cao, Dong, Vemuri, and Du}{Cao
  et~al\mbox{.}}{2020}]%
        {FB-workload}
\bibfield{author}{\bibinfo{person}{Z. Cao}, \bibinfo{person}{S. Dong},
  \bibinfo{person}{S. Vemuri}, {and} \bibinfo{person}{D. Du}.}
  \bibinfo{year}{2020}\natexlab{}.
\newblock \showarticletitle{Characterizing, modeling, and benchmarking RocksDB
  key-value workloads at Facebook}.
\newblock \bibinfo{journal}{\emph{USNIX FAST Conf.}} (\bibinfo{year}{2020}).
\newblock


\bibitem[\protect\citeauthoryear{Cooper, Silberstein, Tam, Ramakrishnan, and
  Sears}{Cooper et~al\mbox{.}}{2010}]%
        {10.1145/1807128.1807152}
\bibfield{author}{\bibinfo{person}{Brian~F. Cooper}, \bibinfo{person}{Adam
  Silberstein}, \bibinfo{person}{Erwin Tam}, \bibinfo{person}{Raghu
  Ramakrishnan}, {and} \bibinfo{person}{Russell Sears}.}
  \bibinfo{year}{2010}\natexlab{}.
\newblock \showarticletitle{Benchmarking Cloud Serving Systems with YCSB}. In
  \bibinfo{booktitle}{\emph{Proceedings of the 1st ACM Symposium on Cloud
  Computing}} (Indianapolis, Indiana, USA) \emph{(\bibinfo{series}{SoCC '10})}.
  \bibinfo{publisher}{Association for Computing Machinery},
  \bibinfo{address}{New York, NY, USA}, \bibinfo{pages}{143–154}.
\newblock
\showISBNx{9781450300360}
\urldef\tempurl%
\url{https://doi.org/10.1145/1807128.1807152}
\showDOI{\tempurl}


\bibitem[\protect\citeauthoryear{Fedorov, Kim, Qin, Gratz, and Reddy}{Fedorov
  et~al\mbox{.}}{2017}]%
        {SPAN}
\bibfield{author}{\bibinfo{person}{Viacheslav Fedorov},
  \bibinfo{person}{Jinchun Kim}, \bibinfo{person}{Mian Qin},
  \bibinfo{person}{Paul~V. Gratz}, {and} \bibinfo{person}{A.~L.~Narasimha
  Reddy}.} \bibinfo{year}{2017}\natexlab{}.
\newblock \showarticletitle{Speculative Paging for Future {NVM} Storage}. In
  \bibinfo{booktitle}{\emph{MEMSYS '17 Proceedings of the International
  Symposium on Memory Systems}}. \bibinfo{address}{Alexandria, Virginia}.
\newblock


\bibitem[\protect\citeauthoryear{Gandhi, Basu, Hill, and Swift}{Gandhi
  et~al\mbox{.}}{2014}]%
        {pagewalker1}
\bibfield{author}{\bibinfo{person}{Jayneel Gandhi}, \bibinfo{person}{Arkaprava
  Basu}, \bibinfo{person}{Mark~D. Hill}, {and} \bibinfo{person}{Michael~M.
  Swift}.} \bibinfo{year}{2014}\natexlab{}.
\newblock \showarticletitle{Efficient memory virtualization: reducing
  dimensionality of nested page walks}.
\newblock \bibinfo{journal}{\emph{IEEE MICRO Conf.}} (\bibinfo{year}{2014}).
\newblock


\bibitem[\protect\citeauthoryear{Gandhi, Karakostas, Ayar, Cristal, Hill,
  McKinley, Nemirovsky, Swift, and Ünsal}{Gandhi et~al\mbox{.}}{2016}]%
        {range-translations}
\bibfield{author}{\bibinfo{person}{J. Gandhi}, \bibinfo{person}{V. Karakostas},
  \bibinfo{person}{F. Ayar}, \bibinfo{person}{A. Cristal}, \bibinfo{person}{MD
  Hill}, \bibinfo{person}{KS McKinley}, \bibinfo{person}{M. Nemirovsky},
  \bibinfo{person}{M. Swift}, {and} \bibinfo{person}{O. Ünsal}.}
  \bibinfo{year}{2016}\natexlab{}.
\newblock \showarticletitle{Range translations for fast virtual memory}.
\newblock \bibinfo{journal}{\emph{IEEE Micro}} (\bibinfo{year}{2016}).
\newblock


\bibitem[\protect\citeauthoryear{Guo, Shan, Luo, Huang, and Zhang}{Guo
  et~al\mbox{.}}{2021}]%
        {guo2021clio}
\bibfield{author}{\bibinfo{person}{Zhiyuan Guo}, \bibinfo{person}{Yizhou Shan},
  \bibinfo{person}{Xuhao Luo}, \bibinfo{person}{Yutong Huang}, {and}
  \bibinfo{person}{Yiying Zhang}.} \bibinfo{year}{2021}\natexlab{}.
\newblock \showarticletitle{Clio: A Hardware-Software Co-Designed Disaggregated
  Memory System}.
\newblock \bibinfo{journal}{\emph{arXiv preprint arXiv:2108.03492}}
  (\bibinfo{year}{2021}).
\newblock


\bibitem[\protect\citeauthoryear{Gupta, Bhattacharyya, Oh, Bhattacharjee,
  Falsafi, and Payer}{Gupta et~al\mbox{.}}{2021}]%
        {10.1109/ISCA52012.2021.00047}
\bibfield{author}{\bibinfo{person}{Siddharth Gupta}, \bibinfo{person}{Atri
  Bhattacharyya}, \bibinfo{person}{Yunho Oh}, \bibinfo{person}{Abhishek
  Bhattacharjee}, \bibinfo{person}{Babak Falsafi}, {and}
  \bibinfo{person}{Mathias Payer}.} \bibinfo{year}{2021}\natexlab{}.
\newblock \bibinfo{booktitle}{\emph{Rebooting Virtual Memory with Midgard}}.
\newblock \bibinfo{publisher}{IEEE Press}, \bibinfo{pages}{512–525}.
\newblock
\showISBNx{9781450390866}
\urldef\tempurl%
\url{https://doi.org/10.1109/ISCA52012.2021.00047}
\showURL{%
\tempurl}


\bibitem[\protect\citeauthoryear{Jin and MA}{Jin and MA}{2000}]%
        {npb}
\bibfield{author}{\bibinfo{person}{H. Jin} {and} \bibinfo{person}{Frumkin MA}.}
  \bibinfo{year}{2000}\natexlab{}.
\newblock \showarticletitle{The OpenMP Implementation of NAS Parallel
  Benchmarks and Its Performance}.
\newblock  (\bibinfo{date}{05} \bibinfo{year}{2000}).
\newblock


\bibitem[\protect\citeauthoryear{Kwon, Fingler, Hunt, Peter, Witchel, and
  Anderson}{Kwon et~al\mbox{.}}{2017}]%
        {strata}
\bibfield{author}{\bibinfo{person}{Youngjin Kwon}, \bibinfo{person}{Henrique
  Fingler}, \bibinfo{person}{Tyler Hunt}, \bibinfo{person}{Simon Peter},
  \bibinfo{person}{Emmett Witchel}, {and} \bibinfo{person}{Thomas Anderson}.}
  \bibinfo{year}{2017}\natexlab{}.
\newblock \showarticletitle{Strata: A Cross Media File System}. In
  \bibinfo{booktitle}{\emph{SOSP '17 Proceedings of the 26th Symposium on
  Operating Systems Principles}}. \bibinfo{publisher}{ACM},
  \bibinfo{address}{Shanghai, China}, \bibinfo{pages}{460 -- 477}.
\newblock


\bibitem[\protect\citeauthoryear{Kwon, Yu, Peter, Rossbach, and Witchel}{Kwon
  et~al\mbox{.}}{2016}]%
        {LargePage-Ingens}
\bibfield{author}{\bibinfo{person}{Y. Kwon}, \bibinfo{person}{H. Yu},
  \bibinfo{person}{S. Peter}, \bibinfo{person}{C.~J. Rossbach}, {and}
  \bibinfo{person}{E. Witchel}.} \bibinfo{year}{2016}\natexlab{}.
\newblock \showarticletitle{Coordinated and Efficient Huge Page Management with
  Ingens}.
\newblock \bibinfo{journal}{\emph{USENIX OSDI}} (\bibinfo{year}{2016}).
\newblock


\bibitem[\protect\citeauthoryear{Lee, Jin, Song, Gong, Bae, Ham, Lee, and
  Jeong}{Lee et~al\mbox{.}}{2020}]%
        {sigarch20}
\bibfield{author}{\bibinfo{person}{Gyusun Lee}, \bibinfo{person}{Wenjing Jin},
  \bibinfo{person}{Wonsuk Song}, \bibinfo{person}{Jeonghun Gong},
  \bibinfo{person}{Jonghyun Bae}, \bibinfo{person}{Tae~Jun Ham},
  \bibinfo{person}{Jae~W. Lee}, {and} \bibinfo{person}{Jinkyu Jeong}.}
  \bibinfo{year}{2020}\natexlab{}.
\newblock \showarticletitle{A Case for Hardware-Based Demand Paging}.
\newblock \bibinfo{journal}{\emph{ACM SIGARCH Architecture Conference}}
  (\bibinfo{year}{2020}).
\newblock


\bibitem[\protect\citeauthoryear{Lesokhin, Eran, Raindel, Shapiro, Grimberg,
  Liss, Ben-Yehuda, Amit, and Tsafrir}{Lesokhin et~al\mbox{.}}{2017}]%
        {10.1145/3037697.3037710}
\bibfield{author}{\bibinfo{person}{Ilya Lesokhin}, \bibinfo{person}{Haggai
  Eran}, \bibinfo{person}{Shachar Raindel}, \bibinfo{person}{Guy Shapiro},
  \bibinfo{person}{Sagi Grimberg}, \bibinfo{person}{Liran Liss},
  \bibinfo{person}{Muli Ben-Yehuda}, \bibinfo{person}{Nadav Amit}, {and}
  \bibinfo{person}{Dan Tsafrir}.} \bibinfo{year}{2017}\natexlab{}.
\newblock \showarticletitle{Page Fault Support for Network Controllers}.
  \bibinfo{publisher}{Association for Computing Machinery},
  \bibinfo{address}{New York, NY, USA}.
\newblock
\showISBNx{9781450344654}
\urldef\tempurl%
\url{https://doi.org/10.1145/3037697.3037710}
\showDOI{\tempurl}


\bibitem[\protect\citeauthoryear{Memaripour and Swanson}{Memaripour and
  Swanson}{2018}]%
        {breeze}
\bibfield{author}{\bibinfo{person}{A. Memaripour} {and} \bibinfo{person}{S.
  Swanson}.} \bibinfo{year}{2018}\natexlab{}.
\newblock \showarticletitle{Breeze: User-Level Access to Non-Volatile Main
  Memories for Legacy Software}.
\newblock \bibinfo{journal}{\emph{IEEE ICDC Conf.}} (\bibinfo{year}{2018}).
\newblock


\bibitem[\protect\citeauthoryear{Navarro, Iyer, Druschel, and Cox}{Navarro
  et~al\mbox{.}}{2002}]%
        {LargePage1}
\bibfield{author}{\bibinfo{person}{J. Navarro}, \bibinfo{person}{S. Iyer},
  \bibinfo{person}{P. Druschel}, {and} \bibinfo{person}{A. Cox}.}
  \bibinfo{year}{Dec. 2002}\natexlab{}.
\newblock \showarticletitle{Practical, Transparent Operating System Support for
  Superpages}.
\newblock \bibinfo{journal}{\emph{ACM SIGOPS Oper. Syst. Rev.}}
  (\bibinfo{year}{Dec. 2002}).
\newblock


\bibitem[\protect\citeauthoryear{Panwar, Bansal, and Gopinath}{Panwar
  et~al\mbox{.}}{2019}]%
        {LargePage2}
\bibfield{author}{\bibinfo{person}{A. Panwar}, \bibinfo{person}{S. Bansal},
  {and} \bibinfo{person}{K. Gopinath}.} \bibinfo{year}{2019}\natexlab{}.
\newblock \showarticletitle{HawkEye: Efficient Finegrained OS Support for Huge
  Pages}.
\newblock \bibinfo{journal}{\emph{ACM ASPLOS}} (\bibinfo{year}{2019}).
\newblock


\bibitem[\protect\citeauthoryear{Park, Cha, Kim, Kwon, Black-Schaffer, and
  Huh}{Park et~al\mbox{.}}{2020}]%
        {perforated-pages}
\bibfield{author}{\bibinfo{person}{C.~H. Park}, \bibinfo{person}{S. Cha},
  \bibinfo{person}{B. Kim}, \bibinfo{person}{Y. Kwon}, \bibinfo{person}{D.
  Black-Schaffer}, {and} \bibinfo{person}{J. Huh}.}
  \bibinfo{year}{2020}\natexlab{}.
\newblock \showarticletitle{Perforated Page: Supporting Fragmented Memory
  Allocation for Large Pages}.
\newblock \bibinfo{journal}{\emph{ACM SIGARCH Conf.}} (\bibinfo{year}{2020}).
\newblock


\bibitem[\protect\citeauthoryear{Pemberton, Kubiatowicz, and Katz}{Pemberton
  et~al\mbox{.}}{[n.\,d.]}]%
        {pemberton2018enabling}
\bibfield{author}{\bibinfo{person}{Nathan Pemberton}, \bibinfo{person}{John~D
  Kubiatowicz}, {and} \bibinfo{person}{Randy~H Katz}.}
  \bibinfo{year}{[n.\,d.]}\natexlab{}.
\newblock \emph{\bibinfo{title}{Enabling efficient and transparent remote
  memory access in disaggregated datacenters}}.
\newblock \bibinfo{thesistype}{Ph.\,D. Dissertation}.
\newblock


\bibitem[\protect\citeauthoryear{Pham, Bhattacharjee, Eckert, and Loh}{Pham
  et~al\mbox{.}}{2014}]%
        {TLB-extend1}
\bibfield{author}{\bibinfo{person}{Binh Pham}, \bibinfo{person}{Abhishek
  Bhattacharjee}, \bibinfo{person}{Yasuko Eckert}, {and}
  \bibinfo{person}{Gabriel Loh}.} \bibinfo{year}{2014}\natexlab{}.
\newblock \showarticletitle{Increasing TLB Reach by Exploiting Clustering in
  Page Translations}.
\newblock \bibinfo{journal}{\emph{IEEE HPCA Conf.}} (\bibinfo{year}{2014}).
\newblock


\bibitem[\protect\citeauthoryear{Shahrad, Balkind, and Wentzlaff}{Shahrad
  et~al\mbox{.}}{2019}]%
        {10.1145/3352460.3358296}
\bibfield{author}{\bibinfo{person}{Mohammad Shahrad}, \bibinfo{person}{Jonathan
  Balkind}, {and} \bibinfo{person}{David Wentzlaff}.}
  \bibinfo{year}{2019}\natexlab{}.
\newblock \showarticletitle{Architectural Implications of Function-as-a-Service
  Computing}. In \bibinfo{booktitle}{\emph{Proceedings of the 52nd Annual
  IEEE/ACM International Symposium on Microarchitecture}} (Columbus, OH, USA)
  \emph{(\bibinfo{series}{MICRO '52})}. \bibinfo{publisher}{Association for
  Computing Machinery}, \bibinfo{address}{New York, NY, USA},
  \bibinfo{pages}{1063–1075}.
\newblock
\showISBNx{9781450369381}
\urldef\tempurl%
\url{https://doi.org/10.1145/3352460.3358296}
\showDOI{\tempurl}


\bibitem[\protect\citeauthoryear{S.Haria, Hill, and Swift}{S.Haria
  et~al\mbox{.}}{2020}]%
        {MOD}
\bibfield{author}{\bibinfo{person}{S.Haria}, \bibinfo{person}{M. Hill}, {and}
  \bibinfo{person}{M. Swift}.} \bibinfo{year}{2020}\natexlab{}.
\newblock \showarticletitle{MOD: Minimally Ordered Durable Datastructures for
  Persistent Memory}.
\newblock \bibinfo{journal}{\emph{Proc. of ACM ASPLOS}} (\bibinfo{year}{2020}).
\newblock


\bibitem[\protect\citeauthoryear{Tramm, Siegel, Islam, and Schulz}{Tramm
  et~al\mbox{.}}{2014}]%
        {Tramm:wy}
\bibfield{author}{\bibinfo{person}{John~R Tramm}, \bibinfo{person}{Andrew~R
  Siegel}, \bibinfo{person}{Tanzima Islam}, {and} \bibinfo{person}{Martin
  Schulz}.} \bibinfo{year}{2014}\natexlab{}.
\newblock \showarticletitle{{XSBench} - The Development and Verification of a
  Performance Abstraction for {M}onte {C}arlo Reactor Analysis}. In
  \bibinfo{booktitle}{\emph{{PHYSOR} 2014 - The Role of Reactor Physics toward
  a Sustainable Future}}. \bibinfo{address}{Kyoto}.
\newblock
\urldef\tempurl%
\url{https://www.mcs.anl.gov/papers/P5064-0114.pdf}
\showURL{%
\tempurl}


\bibitem[\protect\citeauthoryear{Waldspurger}{Waldspurger}{2002}]%
        {vmware}
\bibfield{author}{\bibinfo{person}{Carl~A. Waldspurger}.}
  \bibinfo{year}{2002}\natexlab{}.
\newblock \showarticletitle{Memory Resource Management in VMware ESX Server}.
  In \bibinfo{booktitle}{\emph{OSDI '02 Proceedings of the 5th Symposium on
  Operating Systems Design and Implementation}}. \bibinfo{address}{Boston, MA}.
\newblock


\bibitem[\protect\citeauthoryear{Woo, Ohara, Torrie, Singh, and Gupta}{Woo
  et~al\mbox{.}}{1995}]%
        {SPLASH2}
\bibfield{author}{\bibinfo{person}{Steven~Cameron Woo},
  \bibinfo{person}{Moriyoshi Ohara}, \bibinfo{person}{Evan Torrie},
  \bibinfo{person}{Jaswinder~Pal Singh}, {and} \bibinfo{person}{Anoop Gupta}.}
  \bibinfo{year}{1995}\natexlab{}.
\newblock \showarticletitle{The {SPLASH-2} Programs: Characterization and
  Methodological Considerations}. In \bibinfo{booktitle}{\emph{ISCA '95}}.
  \bibinfo{publisher}{ACM}, \bibinfo{address}{S. Margherita Ligure, Italy},
  \bibinfo{pages}{24--36}.
\newblock


\bibitem[\protect\citeauthoryear{Wu and Reddy}{Wu and Reddy}{2011}]%
        {scmfs}
\bibfield{author}{\bibinfo{person}{Xiaojian Wu} {and}
  \bibinfo{person}{A.~L.~Narasimha Reddy}.} \bibinfo{year}{2011}\natexlab{}.
\newblock \showarticletitle{SCMFS : A File System for Storage Class Memory}. In
  \bibinfo{booktitle}{\emph{SC '11 Proceedings of 2011 International Conference
  for High Performance Computing, Networking, Storage and Analysis}}.
  \bibinfo{publisher}{ACM}, \bibinfo{address}{Seattle, Washington, USA}.
\newblock


\bibitem[\protect\citeauthoryear{Zhan, Bao, Bienia, and Li}{Zhan
  et~al\mbox{.}}{2017}]%
        {10.1145/3053277.3053279}
\bibfield{author}{\bibinfo{person}{Xusheng Zhan}, \bibinfo{person}{Yungang
  Bao}, \bibinfo{person}{Christian Bienia}, {and} \bibinfo{person}{Kai Li}.}
  \bibinfo{year}{2017}\natexlab{}.
\newblock \showarticletitle{PARSEC3.0: A Multicore Benchmark Suite with Network
  Stacks and SPLASH-2X}.
\newblock \bibinfo{journal}{\emph{SIGARCH Comput. Archit. News}}
  \bibinfo{volume}{44}, \bibinfo{number}{5} (\bibinfo{date}{Feb.}
  \bibinfo{year}{2017}), \bibinfo{pages}{1–16}.
\newblock
\showISSN{0163-5964}
\urldef\tempurl%
\url{https://doi.org/10.1145/3053277.3053279}
\showDOI{\tempurl}


\end{thebibliography}
